\documentclass[iop]{emulateapj}

\usepackage{natbib,aas_macros}
\usepackage{multirow,color}

\begin{document}
\shorttitle{
[CII] Halo at $z\sim6$
}
\shortauthors{Fujimoto et al.}
\slugcomment{ApJ in press}

\title{%
First Identification of 10-$\lowercase{\rm kpc}$ \cii\ $158\mu\lowercase{\rm m}$ Halos \\ 
around Star-Forming Galaxies at $\lowercase{z}=5-7$ 
}

\author{%
Seiji Fujimoto\altaffilmark{1,2,3}, 
Masami Ouchi\altaffilmark{1,4}, 
Andrea Ferrara \altaffilmark{5,6}, 
Andrea Pallottini\altaffilmark{5}, 
R. J. Ivison\altaffilmark{7,8}, \\
Christoph Behrens\altaffilmark{5},
Simona Gallerani\altaffilmark{5}, 
{
Shohei Arata\altaffilmark{9}, 
Hidenobu Yajima\altaffilmark{10}, and 
Kentaro Nagamine\altaffilmark{4,9,11} 
}
}

\email{sfseiji@icrr.u-tokyo.ac.jp}

\altaffiltext{1}{%
Institute for Cosmic Ray Research, The University of Tokyo,
Kashiwa, Chiba 277-8582, Japan
}
\altaffiltext{2}{%
Research Institute for Science and Engineering, Waseda University, 3-4-1 Okubo, Shinjuku, Tokyo 169-8555, Japan
}
\altaffiltext{3}{%
National Astronomical Observatory of Japan, 2-21-1, Osawa, Mitaka, Tokyo, Japan
}
\altaffiltext{4}{%
Kavli Institute for the Physics and Mathematics of the Universe (WPI), 
University of Tokyo, Kashiwa 277-8583, Japan
}
\altaffiltext{5}{%
Scuola Normale Superiore, Piazza dei Cavalieri 7, I-56126 Pisa, Italy
}
\altaffiltext{6}{%
Centro Fermi, Museo Storico della Fisica e Centro Studi e Ricerche "Enrico Fermi"
Piazza del Viminale 1, Roma, 00184, Italy
}
\altaffiltext{7}{%
European Southern Observatory, Karl Schwarzschild Str.\ 2, D-85748 Garching, Germany
}
\altaffiltext{8}{%
Institute for Astronomy, University of Edinburgh, Royal Observatory
Blackford Hill, Edinburgh EH9 3HJ, UK
}
\altaffiltext{9}{%
Theoretical Astrophysics, Department of Earth and Space Science, 
Graduate School of Science, Osaka University, Toyonaka, Osaka 560-0043, Japan
}
\altaffiltext{10}{%
Center of Computational Sciences University of Tsukuba, 
Ibaraki 305-8577, Japan}
\altaffiltext{11}{%
Department of Physics \& Astronomy, University of Nevada, 
Las Vegas, 4505 S. Maryland Pkwy, Las Vegas, NV 89154-4002, USA
}

\newcommand{\oiii}{[O\,{\sc iii}]}
\newcommand{\oii}{[O\,{\sc ii}]}
\newcommand{\cii}{[C\,{\sc ii}]}
\newcommand{\ciii}{C\,{\sc iii}]}
\newcommand{\lya}{Ly$\alpha$}
\newcommand{\mum}{$\mu$m}
\newcommand{\dv}{$\Delta v_{\rm Ly\alpha}$}
\newcommand{\ew}{EW$_{\rm 0}$}
\newcommand{\lsun}{$L_{\rm \odot}$}
\newcommand{\msun}{$M_{\rm \odot}$}
\newcommand{\ltir}{$L_{\rm TIR}$}
\newcommand{\nhi}{$N_{\rm HI}$}
\newcommand{\loiii}{$L_{\rm [OIII]}$}
\newcommand{\llya}{$L_{\rm Ly\alpha}$}
\newcommand{\luv}{$L_{\rm UV}$}
\newcommand{\zph}{$z_{\rm ph}$}
\newcommand{\muv}{$M_{\rm UV}$}
\newcommand{\td}{$T_{\rm d}$}
\newcommand{\bd}{$\beta_{\rm d}$}
\newcommand{\md}{$M_{\rm d}$}
\newcommand{\zoiii}{$z_{\rm [OIII]}$}
\newcommand{\zlya}{$z_{\rm Ly\alpha}$}
\newcommand{\zsun}{$Z_{\rm \odot}$}
\newcommand{\mdyn}{$M_{\rm dyn}$}

\def\apj{ApJ}%
\def\apjl{ApJ}%
\def\apjs{ApJS}%

\def\rme{\rm e}
\def\rmFIR{\rm FIR}
\def\itHubble{\it Hubble}
\def\rmyr{\rm yr}

\begin{abstract}
We report the discovery of 10-kpc {\sc [Cii]} 158$\mu$m halos surrounding star-forming galaxies in the early Universe. 
We choose deep ALMA data of 18 galaxies each with a star-formation rate of $\simeq 10-70\,M_\odot$ with no signature of AGN
whose {\sc [Cii]} lines are individually detected at $z=5.153-7.142$,
and conduct stacking of the {\sc [Cii]} lines and dust-continuum 
in the $uv$-visibility plane. 
The radial profiles of the surface brightnesses show a 10-kpc scale {\sc [Cii]} halo 
at the 9.2$\sigma$ level, significantly more extended than the HST stellar continuum data
by a factor of $\sim5$ on the exponential-profile basis, as well as the dust continuum.
We compare the radial profiles of {\sc [Cii]} and Ly$\alpha$ halos
universally found in star-forming galaxies at this epoch, and find that the scale lengths agree within $1\sigma$ level.
While two independent hydrodynamical zoom-in simulations match the dust and stellar continuum properties, 
the simulations cannot reproduce the extended \cii\ line emission.
The existence of the extended {\sc [Cii]} halo is the evidence of outflow remnants in the early galaxies 
and suggest that the outflows may be dominated by cold-mode outflows expelling the neutral gas. 
\end{abstract}

\keywords{%
galaxies: formation ---
galaxies: evolution ---
galaxies: high-redshift 
}

\section{Introduction}
\label{sec:intro}
Galaxy size and morphological studies in the early Universe provide important insights into the initial stage of galaxy formation and evolution.  
The size and morphology in the rest-frame ultra-violet (UV) and far-infrared (FIR) wavelengths trace the areas of young star formation and the active starbursts 
that are less and heavily obscured by dust, respectively. 
The {\sc [Cii]} ${\rm ^{2}P_{3/2}} \rightarrow \rm ^{2}{\rm P_{1/2}} $ fine-structure transition at 1900.5469 GHz (157.74 $\mu$m) is a dominant coolant of the inter-stellar medium (ISM) in galaxies \citep[e.g.,][]{stacey1991,delooze2014}, 
whose size and morphology are strong probes of ISM properties. 
Comparing the size and morphology in the rest-frame UV+FIR continuum and the {\sc [Cii]} 158-$\mu$m line is thus important to comprehensively understand the evolutionary process via the star-formation surrounded by the ISM. 

The {\it Hubble\,Space\,Telescope} (HST) has revealed the size and morphological properties in the rest-frame UV wavelengths for the high-redshift galaxies up to $z\sim10$ \citep[e.g.,][]{oesch2010, ono2013, shibuya2015, bouwens2017, kawamata2018}. 
These HST studies show that star-forming galaxies generally have an exponential-disk profile and become compact toward high redshifts. 

The Atacama Large Millimeter / submillimeter Array (ALMA) has opened our views to the obscured star-formation and the {\sc [Cii]} line properties in the rest-frame FIR wavelength up to $z\sim7$ \citep[e.g.,][]{watson2015,maiolino2015,capak2015,pentericci2016,knudsen2016,matthee2017,carniani2018b,smit2018,hashimoto2018b}. 
There have also been several attempts to measure the size and morphology in the rest-frame FIR continuum and the {\sc [Cii]} line for such high redshift galaxies at $z\sim$ 5$-$7, 
where \cite{carniani2017} report that the effective radius of the {\sc [Cii]} line-emitting region is larger than that of the rest-frame UV region. 
However, large uncertainties still remain due to the small number statistics and observational challenges. 

One critical challenge is sensitivity. 
The recent ALMA studies show that signal-to-noise ratio (S/N) $>$ 10 is needed to obtain reliable size measurement results both on the image-based and visibility-based analyses \citep[e.g.,][]{simpson2015a,ikarashi2015}, 
while the majority of the previous ALMA detections of the dust continuum and the {\sc [Cii]} line from $z\sim$ 5$-$7 galaxies show the S/N less than 10. 
If the S/N level is poor, noise fluctuations significantly affect the profile fitting results. 
Moreover, \cite{hodge2016} show that the combination of the original smoothed galaxy profile and the noise fluctuations can make the morphology more clumpy. 
To obtain the reliable size and morphological results, extensively deep observations are thus required. 

In this paper, we determine the size and morphology for the dust continuum and the {\sc [Cii]} line in the star-forming galaxies at $z=5-7$ via the stacking technique in the $uv$-visibility plane, 
utilizing new and archival deep ALMA Band 6/7 data. 
In conjunction with deep HST images, we study the general morphology of the total star-formation and the ISM in the epoch of re-ionization. 
The structure of this paper is as follows. 
In Section 2, the observations and the data reduction are described.  
Section 3 outlines the method of {\sc [Cii]} line detections, line velocity width, source position measurements, and the stacking processes of ALMA and HST data.  
We report the results of the radial profiles of the {\sc [Cii]} line, rest-frame FIR, and rest-frame UV wavelengths in Section 4. 
In Section 5, we discuss the physical origin of the extended {\sc [Cii]} line emission, comparing with the zoom-in cosmological simulation results. 
A summary of this study is presented in Section 6. 

\renewcommand{\thefootnote}{\fnsymbol{footnote}}

\begin{table*}
\caption{Our ALMA Sample 
\label{tab:our_alma}}
\begin{tabular}{lcccccccccc}
\hline
\hline
 Target & R.A & Dec. & $z_{{\rm {[CII]}}}$ ($z_{{\rm Ly}\alpha}$) & $M_{\rm UV}$ & EW$_{\rm Ly\alpha}$ &$\sigma_{\rm cont.}$$^{\dagger}$ & Beam & ALMA ID & HST & Ref. \\
            &  (J2000) & (J2000)    &                                                                       &           (mag)     &     (${\rm \AA}$)    &  ($\mu$Jy/beam)  & ($''\times''$)  &                 &  &         \\
            &        &           &              (1)                                                     &          (2)          &            (3)              &         (4)               &        (5)          &  (6)           & (7)  & (8)   \\ \hline 
\multicolumn{11}{c}{Literature} \\ \hline
WMH5            & 36.612542         & $-$4.877333 & 6.069 (6.076)  & $-$22.6 & 13.0 & 8   & 0.50$\times$0.46       & 2013.1.00815.S     & N   & W15, J17 \\
                &                  &           &                 & & &      &            & 2015.1.00834.S     &    & W15, J17 \\
CLM1            & 37.012319        & $-$4.271706 & 6.166 (6.176)  & $-$22.6 & 50.0 & 18   & 0.52$\times$0.45      &  2013.1.00815.S    & N   & W15 \\
COS301855       & 150.125803       & 2.266613 & 6.854 ($$-$$) & $-$21.9 & $-$2.9 & 27   & 1.08$\times$0.74           &  2015.1.01111.S    & Y   & S18 (S15) \\
COS298703       & 150.124400       & 2.217294 & 6.808 (6.816) & $-$22.0 & 16.2 & 25 &1.07$\times$0.74            &  2015.1.01111.S    & Y   & S18 (S15)\\
NTTDF6345       & 181.403878       & $-$7.756192 & 6.698 (6.701) & $-$21.5 & 15.0 & 20& 1.25$\times$0.97           &  2015.1.01105.S  & N   & P16 \\
BDF2203         & 336.958267      & $-$35.147529 & 6.122 (6.118) &  $-$20.9 & 9.9  & 20 & 1.85$\times$1.05         & 2016.1.01240.S   & N$^{\dagger\dagger}$ & C18 \\      
COS13679        & 150.099014      & 2.343517           & 7.142 (7.145) & $-$21.4 & 15.0  & 18 & 0.85$\times$0.85 &  2015.1.01105.S    & N$^{\ddagger}$   & P16 \\
COS24108        & 150.197356      & 2.478931           & 6.623 (6.629) & $-$21.6 & 27.0  & 20 & 0.81$\times$0.75 & 2015.1.01105.S     & N$^{\ddagger}$   & P16 \\
Hz1  		    & 149.971828      &  2.118142           & 5.689 (5.690) & $-$22.0 & 5.3 & 27& 0.75$\times$0.52   & 2012.1.00523.S     & Y   & C15 (B17)\\
Hz2  		    & 150.517186      & 1.928936           & 5.670 (5.670) & $-$21.9 & 6.9 & 35& 0.83$\times$0.53    & 2012.1.00523.S     & N$^{\ddagger}$   & C15 (B17)\\
Hz3  		    & 150.039247      & 2.3371611         & 5.542 (5.546) & $-$21.7 & $-$3.6 & 47& 0.77$\times$0.42    & 2012.1.00523.S   & Y   & C15 (B17)\\
Hz4  		    & 149.618760      & 2.051850          & 5.544 (5.310) & $-$22.3 & 10.2 & 64& 0.89$\times$0.51   & 2012.1.00523.S      & Y   & C15 (B17)\\
Hz6  			& 150.089576     & 2.586324          & 5.293 (5.290)& $-$22.8 & 8.0 & 32& 0.67$\times$0.50      & 2012.1.00523.S      & Y   & C15 (B17)\\
Hz7  			& 149.876925     & 2.134061 & 5.253 (5.250) & $-$21.8 & 9.8  & 35& 0.47$\times$0.38 &  2012.1.00523.S   & Y & C15 (B17)\\
Hz8  			& 150.016894     & 2.626631 & 5.153 (5.148) & $-$21.8 & 27.1 & 30& 0.40$\times$0.29 &  2012.1.00523.S   & Y & C15 (B17)\\
Hz9  			& 149.965404     & 2.378358 & 5.541 (5.548) & $-$21.9 & 14.4 & 43& 0.64$\times$0.54 &  2012.1.00523.S   & Y & C15 (B17)\\ \hline
\multicolumn{10}{c}{New Detection} \\ \hline
NB816$-$S$-$61269 & 34.438567       & $-$5.493392 & 5.684 (5.688) & $-$20.4 & 93.3 & 22 & 0.45$\times$0.42 &  2012.1.00602.S & N &  F16\\
WMH13             & 149.985580  & 2.207528 & 5.985 (5.983) & $-$22.0 & 27.0 & 16 & 1.15$\times$0.89 & 2013.1.00815.S  & N & W15\\
\hline
\end{tabular}
\footnotesize{Notes: 
(1) Spectroscopic redshift determined by the {\sc [Cii]} (Ly$\alpha$) line emission. 
(2) Absolute magnitudes. 
(3) Rest-frame Ly$\alpha$ EW. 
(4) One sigma noise measured by the standard deviation of the pixel values in the continuum map before primary beam correction. 
(5) Synthesized beam size of our ALMA maps (weighting = "natural"). 
(6) ALMA project ID.  
(7) "Y" ("N") indicates the sources (not) included in the ALMA-HST sample. 
(8) ALMA (HST) data reference (W15: \citealt{willott2015}, J17: \citealt{jones2017}, S18: \citealt{smit2018}, P16: \citealt{pentericci2016}, C18: \citealt{carniani2018b}, S15: \citealt{smit2015}, C15: \citealt{capak2015}, B17: \citealt{barisic2017}, F16: \citealt{fujimoto2016}). 
}\\
$\dagger$ Our additional flagging and difference in the imaging parameter setting may produce different values from the data references. \\
$\dagger\dagger$ Although there is the F105W data, we do not include this source in the ALMA-HST sample due to the differences in the PSF and the rest-frame wavelength from the F160W data. \\ 
$\ddagger$ Although there is the F160W data, we do not include these sources in the ALMA-HST sample due to the large offsets even after the astrometry correction (see text). \\  
\vspace{0.2cm}
\end{table*}


Throughout this paper, we assume a flat universe with 
$\Omega_{\rm m} = 0.3$, 
$\Omega_\Lambda = 0.7$, 
$\sigma_8 = 0.8$, 
and $H_0 = 70$ km s$^{-1}$ Mpc$^{-1}$. 
We use magnitudes in the AB system \citep{oke1983}. 

\section{Sample and Data Reduction} 
\label{sec:data}

\subsection{Our ALMA Sample}
\label{sec:sample}

The sample is drawn mainly from the literature \citep{capak2015,willott2015,pentericci2016,smit2018,carniani2018b,jones2017},
selecting only star-forming galaxies at $z>5$ whose \cii\ lines have been detected (at signal-to-noise, S/N $\gtrsim$ 5) with ALMA. 
To obtain reliable results for representative galaxies in the early Universe, 
we limit our sample to galaxies with 
(i) star-formation rates (SFRs), $< 100$\,M$_{\odot}$\,yr$^{-1}$, 
(ii) no indication of AGN activity,
(iii) no giant Ly$\alpha$ systems, such as Himiko \citep{ouchi2009} and CR7 \citep{matthee2015}, 
(iv) no signs of gravitationally lensing, e.g. galaxies behind massive galaxy clusters, 
(v) \cii\ line emission with a full width at half maximum (FWHM) broader than 80\,km s$^{-1}$, 
and (vi) \cii\ line detections that are reproduced in our own data reduction.  
We adopt (v) because the thermal noise fluctuation can produce peaky false source signals even with S/N $>$ 5, 
when we examine the large volume data such as the ALMA 3D data cubes. 
Note that our sample does not include the tentative \cii\ line detections reported 
in the ALMA blind line survey \citep{aravena2016, hayatsu2017}, 
because these tentative \cii\ detections have not been spectroscopically confirmed.
We identify 16 \cii\ line sources that meet the above criteria in the literature. 
Table \ref{tab:our_alma} summarises our sample and the references that describe the relevant ALMA observations.

\begin{figure}[h]
\begin{center}
\includegraphics[trim=0.2cm 0.4cm 0.2cm 0.4cm, clip, angle=90,width=0.5\textwidth]{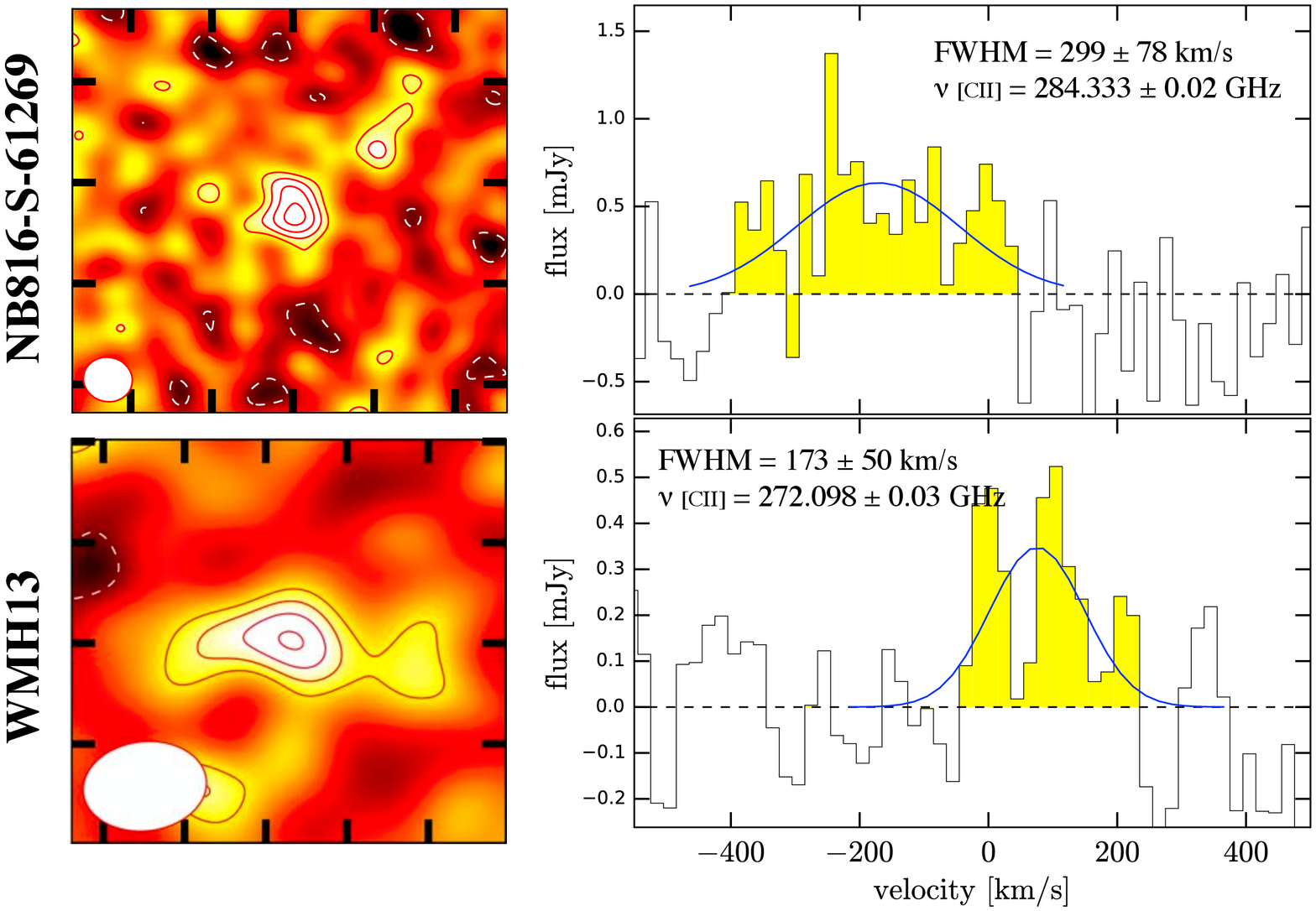}
 \caption[]{
New {\sc [Cii]} line detections of NB816-S-61269 (top) and WMH13 (botttom).  
{\it \bf Left:} Natural-weighted $4''\times4''$ field image of the velocity-integrated \cii\ line intensity (moment zero)
with contours at the $-2\sigma$ (white), 2$\sigma$, 3$\sigma$, 4$\sigma$, and 5$\sigma$ (red) levels.  
The ALMA synthesized beams are presented at the bottom left. 
{\it \bf Right:} \cii\ line spectra with an aperture diameter of $1\farcs2$. 
The solid curves denote the best-fit profile of the single Gaussian with the best-fit values of the FWHM and the frequency peak. 
The yellow shades present the integrated velocity ranges for the \cii\ line intensity maps in the left panel. 
The velocities are relative to the Ly$\alpha$ line obtained in previous studies \citep{ouchi2010,willott2013b}.
\label{fig:new_cii}}
\end{center}
\end{figure}

In addition to the literature sample, 
we include new \cii\ line detections of two star-forming galaxies,
NB816-S-61269 \citep{ouchi2008,fujimoto2016} and WMH13 \citep{willott2013}
at $z=5.688$ and $5.983$, respectively.  In Figure \ref{fig:new_cii}, 
we present the velocity-integrated maps and the spectra for these two
\cii\ detections. In the velocity-integrated maps of NB816-S-61269
and WMH13, the \cii\ line is detected with peak S/N levels of 5.6 
and 5.2, respectively; rest-frame FIR dust continuum emission is not
detected from either galaxy. The details of the ALMA observations for
these additional sources are listed in Table \ref{tab:our_alma}.

From the literature and the additional samples, we obtain a total of 18 {\sc [Cii]} line sources. 
The 18 {\sc [Cii]} line sources have the spectroscopic redshifts determined by the {\sc [Cii]} lines ($z_{\rm [CII]}$) and the absolute rest-frame UV magnitudes ($M_{\rm UV}$) in the ranges of $z_{\rm [CII]}=5.153-7.142$ and $M_{\rm UV}\simeq$ $-$22.8 to $-$20.4 (SFR $\simeq10-70\,M_\odot$/yr). 
We summarize the physical properties of $z_{\rm [CII]}$, $M_{\rm UV}$, and the Ly$\alpha$ equivalent-width (EW$_{\rm Ly\alpha}$) in Table 1. 

\subsection{ALMA Data}
\label{sec:alma_data}

We reduce the ALMA data for our sample with the Common Astronomy Software Applications package (CASA; \citealt{mcmullin2007})   
in the standard manner with the scripts provided by the ALMA observatory. 
In this process, we carry out re-calibrations for the flux density and additional flagging for bad antennae if we find problems in the final images that shows striped patterns and/or significantly higher noise levels than expected. 
The continuum images and line cubes are produced by the CLEAN algorithm with the {\sc tclean} task with a pixel scale of $0\farcs01$. 
For the line cubes, the velocity channel width is re-binned to 20\,km\,s$^{-1}$, where the velocity center is adjusted to the Ly$\alpha$ redshift. 
We do not CLEAN the line cubes because the {\sc [Cii]} line is faint in each 20-km\,s$^{-1}$ channel. 
The CLEAN boxes were set at the peak pixel positions with S/N $\geq$ 5 in the auto mode, 
and the CLEAN routines were proceeded down to the 3$\sigma$ level. 
We list the standard deviation of the pixel values in a final natural-weighted image and a synthesized beam size for the continuum image in Table 1. 

Note that the continuum is subtracted from the $uv$-data of the line cubes 
for 4 sources (Hz4, Hz6, Hz9, and WMH5) whose continuum emission is individually detected \citep{capak2015,willott2015}. 
The continuum level is estimated from the channels in the velocity range of $v>$ $|$2 $\times$ FWHM$|$ in the same baseband as the \cii\ line emission. 

\subsection{HST Data}
\label{hst_data}
To study the rest-frame UV properties of our sample, 
we also use the HST Wide Field Camera 3 (WFC3) in F160W, 1.54 $\mu$m ($H$-band), images from the Hubble Legacy Archive,
where we obtain final flat-field and flux-calibrated science products. 

To correct the potential offsets of the HST astrometry \citep[e.g.,][]{rujopakarn2016,dunlop2017}, 
we calibrate the astrometry of the $H$-band maps with the Gaia Data Release 2 catalog \citep{gaia2018}.  
First, we identify bright objects in the $H$-band images with {\sc sextractor} version 2.5.0 \citep{bertin1996}. 
Second, we cross-match the bright $H$-band objects and the GAIA catalog.  
Finally, we evaluate offsets between the bright $H$-band object centers and the GAIA catalog positions. 
We find that the bright $H$-band object centers indeed have the offsets from the GAIA catalog 
in the range of $\sim0\farcs1$$-$$0\farcs3$. 
We correct the astrometry of each $H$-band map to match the GAIA catalog based on these offsets. 
With the above procedure, 
the majority of our sample shows that the \cii\ line and the $H$-band continuum have a consistent peak position within the offset smaller than $\sim0.''1$. 
However, the large offset over $0.''5$ still remains in some cases, 
probably because the astrometry correction does not work successfully, 
or we witness the intrinsic offset between the \cii\ line and the rest-frame UV continuum \citep[e.g.,][]{maiolino2015}. 
In any cases, these objects with the large offsets cause the smearing effect in the stacking results. 
To securely study the morphological property from the stacking results, 
we do not include these objects in the following HST data analyses. 
We identify that 9 out of 18 sources in our sample 
have been observed with the HST/$H$-band whose astrometry is successfully corrected. 
We refer to the 9 and the 18 sources as the "ALMA-HST" and "ALMA-ALL" samples, respectively. 
In Table 1, we summarize the HST data references and the sources included in the ALMA-HST sample. 

Note that we confirm that the ALMA astrometry is well consistent with the GAIA catalog within a milli-arcsec scale 
via the bright quasars used as the phase calibrators in the ALMA observations. 
Thus, we do not carry out any astrometry corrections for our ALMA maps.

\section{Data Analysis}
\label{sec:data_analysis}

\begin{figure}[h]
\begin{center}
\includegraphics[trim=0cm 0cm 0cm 0cm, clip, angle=0,width=0.5\textwidth]{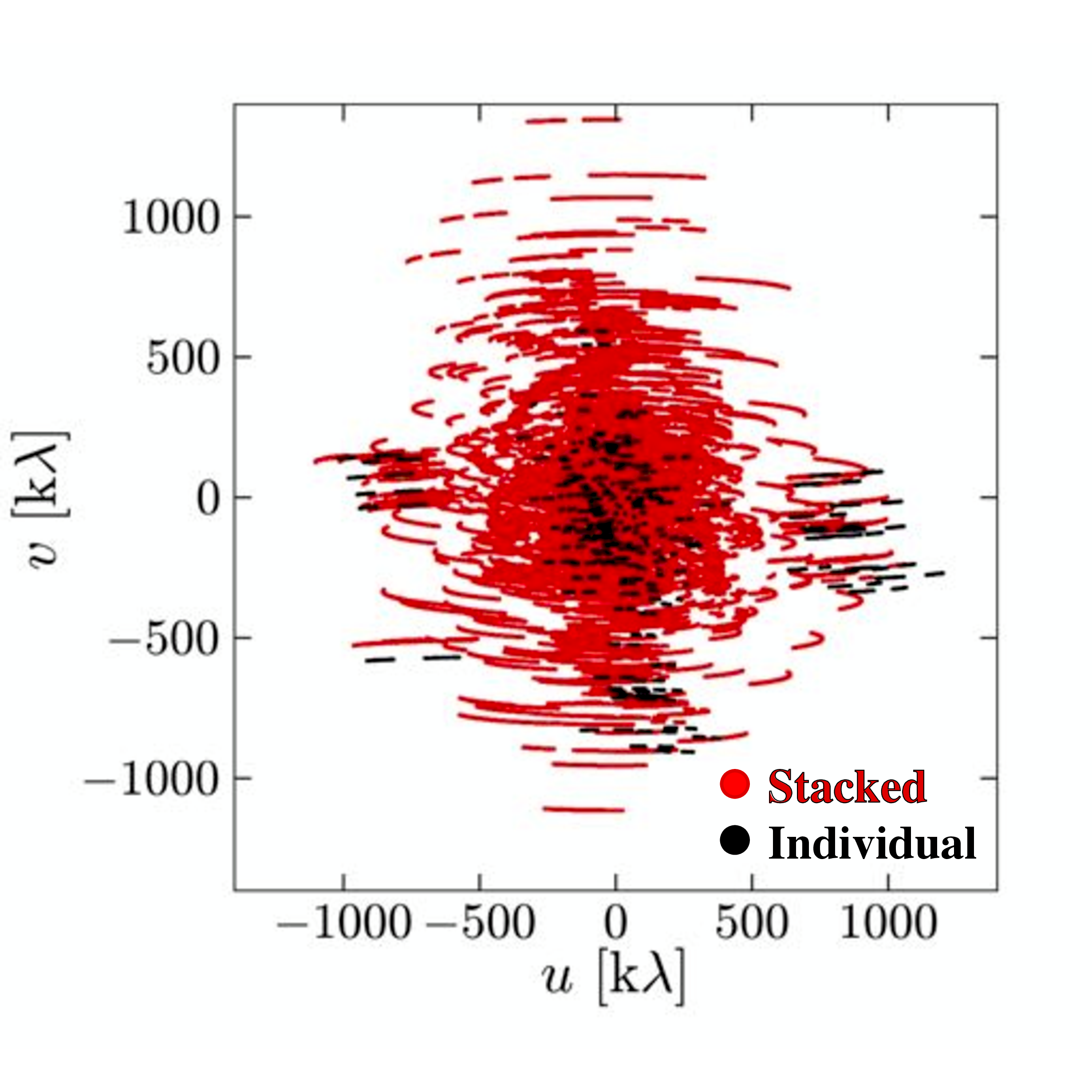}
 \vspace{-0.8cm}
 \caption[]{
$uv$-visibility coverage for individual and stacked data. 
For the individual data, we present Hz3 data as an example. 
For the stacked data, the $uv$-visibility coverage less than 500 k$\lambda$ is well sampled in circular symmetrically, 
which enables us to investigate the diffuse, extended structures.
\label{fig:uv-plane}}
\end{center}
\end{figure}

\begin{figure*}[t]
\begin{center}
\includegraphics[trim=0cm 1.4cm 0cm 0cm, clip, angle=0,width=1.0\textwidth]{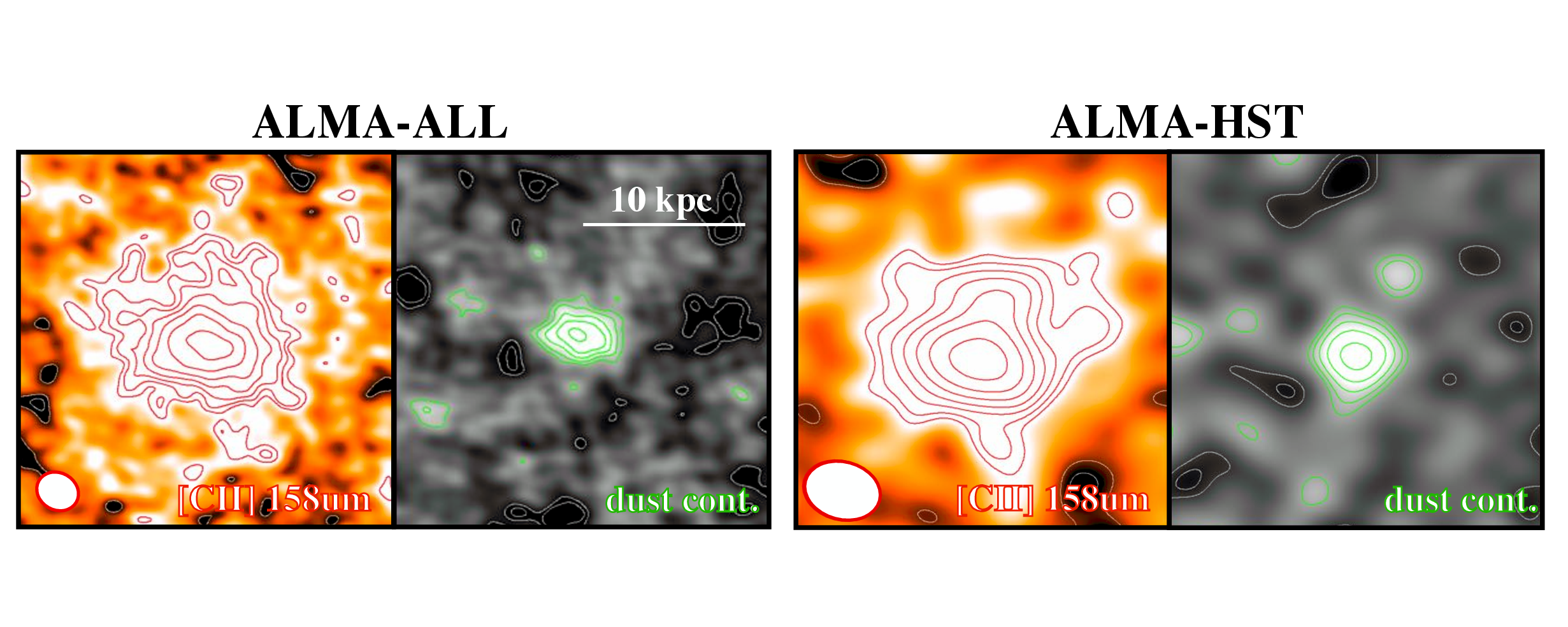}
\vspace{0.2cm}
 \caption[]{
Natural-weighted $4''\times4$ field image after the visibility-based stacking of the {\sc [Cii]} line and the dust continuum 
for the ALMA-ALL (left) and ALMA-HST (right) samples.  
The red and green contours denote the 2, 2$\sqrt{2}$, 4, ... $\times\sigma$ levels of the {\sc [Cii]} line and the dust continuum emission, 
while the white contours indicate the $-2\sigma$ and $-2\sqrt{2}$$\sigma$ levels.  
The synthesized beams are presented at the bottom left in each panel. 
\label{fig:stack_image}}
\end{center}
\end{figure*}

\subsection{3D Position in ALMA Cube}
\label{sec:lp_measure}
To carry out stacking for the \cii\ line and the rest-frame FIR continuum, 
we estimate source centroids for the 18 \cii\ line sources in the ALMA 3-dimensional data cubes via the following six steps: 
(1) We create fiducial \cii\ velocity-integrated maps in the velocity range, 
$\sim$100--900\,km\,s$^{-1}$, that maximizes the S/N level of the \cii\ line detection.
(2) We measure fiducial positional centroids based on the peak pixel positions (pixel scale = $0\farcs01$)
in the fiducial \cii\ velocity-integrated maps, having smoothed spatially with a $uv$-taper of $0\farcs6$.  
(3) We produce \cii\ spectra with an aperture diameter of $1.''2$ at the fiducial source centroids. 
(4) We obtain the peak frequencies and FWHMs of the \cii\ line emission by fitting a single Gaussian to the \cii\ spectra.
(5) We re-create velocity-integrated maps with velocity ranges of $2\times$ the FWHM. 
(6) We measure final positional centroids in the new velocity-integrated map in the same manner as step (2). 
Note that we use the smoothed map (via the $uv$-taper) instead of the naturally-weighted map in steps 
(2) and (6) because Monte-Carlo simulations in the $uv$-visibility plane show that smoothed maps have lower 
uncertainties in the positional measurements than the intrinsic maps \citep{fujimoto2018}. 
We list the final positional centroids and redshifts in Table 1.

\subsection{ALMA Visibility-based Stacking}
\label{sec:alma_stack} 

We carry out visibility-based stacking for our ALMA data via the following procedure. 
First, we split the visibility data into the \cii\ line and the rest-frame FIR continuum datasets. 
For the \cii\ line dataset, we extract the visibility data with the \cii\ line channels across a 
velocity range of 100\,km\,s$^{-1}$ (= $\pm$ 50\,km\,s$^{-1}$), 
where the velocity center is the \cii\ frequency peak (the 3D position in our ALMA cubes).  
We do not adopt a wider velocity range because of the potential contamination of 
the close companions \citep{jones2017,carniani2018b}. 
For the rest-frame FIR continuum dataset, we produce the visibility data whose \cii\ line 
channels in a velocity range of 2$\times$FWHM are fully removed.  
Second, we shift the coordinate of the visibility datasets by re-writing the source center 
determined in Section \ref{sec:lp_measure}
as "00:00:00.00 00:00:00.0" with {\sc stacker} \citep{lindroos2015}.  
Third, we combine the visibility datasets with the {\sc concat} task. 
Fourth, we re-calculate the data weights for the combined visibility datasets with the 
{\sc statwt} task, based on the scatter of visibilities, which includes the effects
of integration time, channel width and system temperature.
Finally, we obtain the stacked datasets of the \cii\ line and the rest-frame FIR continuum. 
The central frequency in the \cii\ line dataset is 271.167 GHz 
which corresponds to the \cii\ redshift at $z=6.01$. 
Assuming the redshift of $z=6.01$ as the weighted average source redshift of our sample, 
we adopt the angular scale of $1''=5.7$ kpc in the following analyses. 
Note that 
we adopt the $H$-band peak positions (Section \ref{sec:hst_stack}) 
as the common stacking center for the ALMA-HST sample. 

Figure \ref{fig:uv-plane} indicates the $uv$-visibility coverage after the
visibility-based stacking for the ALMA datasets of the ALMA-ALL sample. 
For comparison, the $uv$-visibility coverage for an individual dataset, before stacking, 
is also plotted. In the stacked data, the $uv$-visibility coverage is well sampled, 
especially for the short baselines, $<500$\,k$\lambda$, which is important to recover the
flux density from diffuse, extended structures.  

In Figure \ref{fig:stack_image}, we show the natural-weighted images of the {\sc [Cii]} line 
and dust continuum after the visibility-based stacking for the ALMA-ALL (ALMA-HST) sample, 
where the standard deviation of the pixel values in the dust continuum image achieves 4.1 (8.3) $\mu$Jy/beam 
with the synthesized beam size of $0\farcs43\times0\farcs36$ ($0\farcs74\times0.58$). 
The peak pixel signal-to-noise (S/N) ratio shows 21$\sigma$ (20$\sigma$) and 10$\sigma$ (8$\sigma$) significance levels 
for the {\sc [Cii]} line and dust continuum, respectively, for the ALMA-ALL (ALMA-HST) sample. 
The spatially resolved \cii\ line emission in the ALMA-ALL sample is detected at the 9.3 $\sigma$ level in the aperture radius of 10 kpc 
even after masking the emission in a central area up to 2$\times$FWHM of the ALMA synthesized beam, based on the random-aperture method. 
Because the extended structure is difficult to be modeled by the {\sc clean} algorithm perfectly, 
we use the dirty images for both the \cii\ line and the rest-frame FIR continuum in the following analyses.

\begin{figure*}[t]
\begin{center}
\includegraphics[trim=0cm 0cm 0cm 0.2cm, clip, angle=0,width=.95\textwidth]{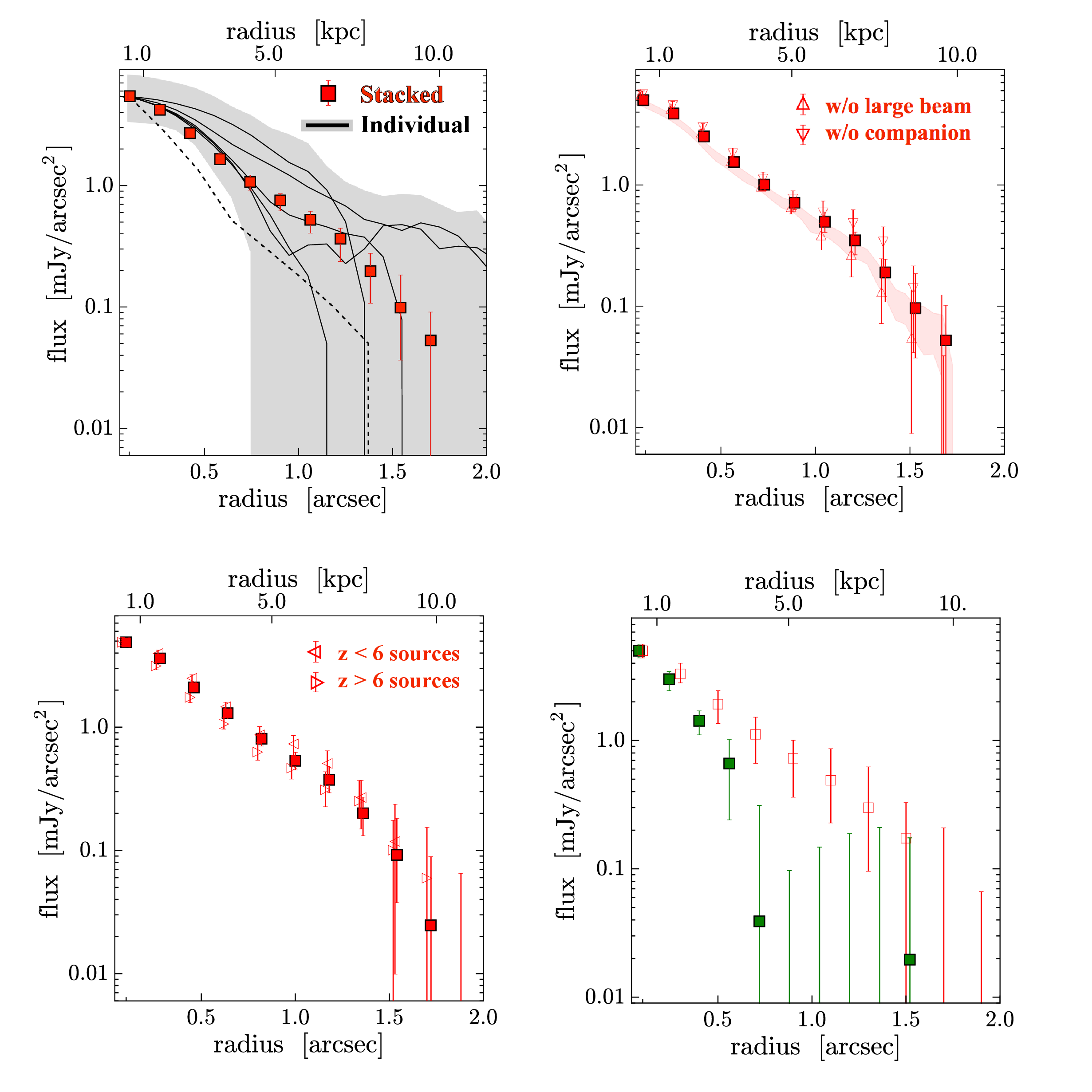}
 \vspace{-0.6cm}
 \caption[]{
Radial surface brightness profile of the \cii\ line (red filled squares) and dust-continuum (green filled squares) emission for the ALMA-ALL sample. 
{\it \bf Top Left:} 
The black solid curves denote the individual results from five \cii\ line sources 
whose \cii\ lines are detected with high S/N levels and with the ALMA beam sizes of $\gtrsim0\farcs8$ to recover the diffuse, extended structures.
The gray shades indicate the error range of the individual results.
The black dashed curve presents the synthesized ALMA beam in the ALMA-ALL sample. 
{\it \bf Top Right:} 
The red shade shows the 16--84 percentile of the sample variance (see text). 
The open symbols indicate the re-stacked results without the \cii\ line sources that are 
I) taken with the lowest resolutions (BDF2203, NTTDF6345, and WMH13; upward triangle), and 
II) reported to have companions (WMH5, Hz2, Hz6, and Hz8; downward triangle).   
{\it \bf Bottom Left:} 
The re-stacked results for the low- ($z<6$; leftward triangle) and high- ($z>6$; rightward triangle) 
redshift subsamples among the 18 \cii\ line sources. 
{\it \bf Bottom Right:}
whose peak S/N ratio is reduced down to the level comparable to the dust continuum map.
All radial profiles are normalized to the peak value of the \cii\ line.
\label{fig:radial_ind}}
\end{center}
\end{figure*}

In Figure \ref{fig:radial_ind}, we present the radial surface brightness profile of the stacked \cii\ line, 
and summarize various tests for the extended \cii\ line structure. 
First, 
we compare our stacking and individual results. 
In the top left panel of Figure \ref{fig:radial_ind}, we show the individual results for several \cii\ line sources  
whose lines are detected at high S/N, with an ALMA beam size of $\gtrsim0\farcs8$ to
recover the diffuse, extended structures.  
We find that the stacked results are consistent with the individual results within the scatter,
suggesting that our ALMA stacking result provides a faithful representative of the 18 \cii\ line sources.
Second,
we evaluate the uncertainty of the sample variance. 
We make 18 newly stacked data with 17 \cii\ line sources, i.e., in each newly stacked data we remove 
one source from the full sample, and derive the 18 \cii\ radial profiles. 
In the top right panel of Figure \ref{fig:radial_ind}, 
the red shaded area indicates the 16--84 percentiles of these 18 radial profiles. 
The \cii\ radial profile is extended up to the radius of $\sim$ 10 kpc even including the sample variance, 
suggesting that the sample variance does not change our results of the existence of the extended \cii\ line emission. 
Third, 
we investigate whether the extended \cii\ line structure is caused by any specific data properties. 
We remove the sources that are 
I) taken with the lowest resolutions (BDF2203, NTTDF6345, and WMH13), 
and II) reported to have companions (WMH5, Hz2, Hz6, and Hz8; \citealt{jones2017,carniani2018b}), 
and obtain newly stacked data. 
In the top right panel of Figure \ref{fig:radial_ind},
we present the radial profiles of the \cii\ line emission in the newly stacked data. 
We find that the newly stacked \cii\ line profiles reproduce the extended structures 
that are well consistent with the original stacking result in the ALMA-ALL sample. 
This indicates that the extended \cii\ line structure is not caused either by 
the bias to the low-resolution data or the contamination of the companions. 
Fourth, 
we examine the surface brightness dimming effect among our sample. 
We divide the 18 \cii\ line sources into two subsamples: low ($z<6$) and high-redshift ($z>6$) samples, 
and obtain other newly stacked data. 
In the bottom left panel of Figure \ref{fig:radial_ind},
we show the radial profiles of the \cii\ line emission in both subsamples.  
We find that the \cii\ line profiles in both subsamples reproduce the extended structures 
that have good agreements with the original stacking result in the ALMA-ALL sample. 
This suggests that the surface brightness dimming effect does not significantly affect our stacking results. 
Fifth, we compare the structures of the \cii\ line and the dust continuum in the same significance level. 
We produce a random noise map smoothed by the ALMA beam, and combine the noise and the stacked \cii\ line maps. 
Changing the noise levels, we obtain the noise-enhanced \cii\ line map whose peak S/N ratio becomes comparable to the dust continuum one. 
We create the 50 noise-enhanced \cii\ line maps. 
In the bottom right panel of Figure \ref{fig:radial_ind}, 
we show the 16--84th percentile of the \cii\ radial profile in the noise-enhanced maps.  
We find that the \cii\ line profile still exceeds more than the dust continuum in these noise-enhanced maps, 
showing that the different structures between the \cii\ line and the dust continuum 
are not mimicked by the difference in the dynamic range.

\subsection{HST/H-band Stacking}
\label{sec:hst_stack}  

We have performed image-based stacking for the ALMA-HST sample, 
exploiting their deep archival HST $H$-band imaging. 
Before stacking, we carry out the following procedure: 
1) We cut out $8''\times8''$ stamps from the $H$-band images, 
around the \cii\ line sources, and set the pixel scale to $0\farcs01$, 
which corresponds to our ALMA images. 
2) We identify low-redshift contaminants within $2\farcs0$ from the \cii\ sources, 
by cross-matching the \cii\ line source positions with photometric redshift catalogs \citep{ilbert2013,skelton2014}.  
3) We remove the low-redshift contaminants from the $H$-band images by fitting 
S$\acute{\rm e}$rsic profiles \citep{sersic1963} with {\sc galfit} \citep{peng2010}.  
We then proceed to generate an average stack, weighted by the noise levels of
the ALMA images of the \cii\ line source. 
This is because the visibility-based stacking for our ALMA data is weighted by the
visibility scatter, which generally corresponded to the noise levels on the ALMA images.

\begin{figure*}
\begin{center}
\includegraphics[trim=0.cm 0cm 0cm 0.5cm, clip, angle=0,width=0.95\textwidth]{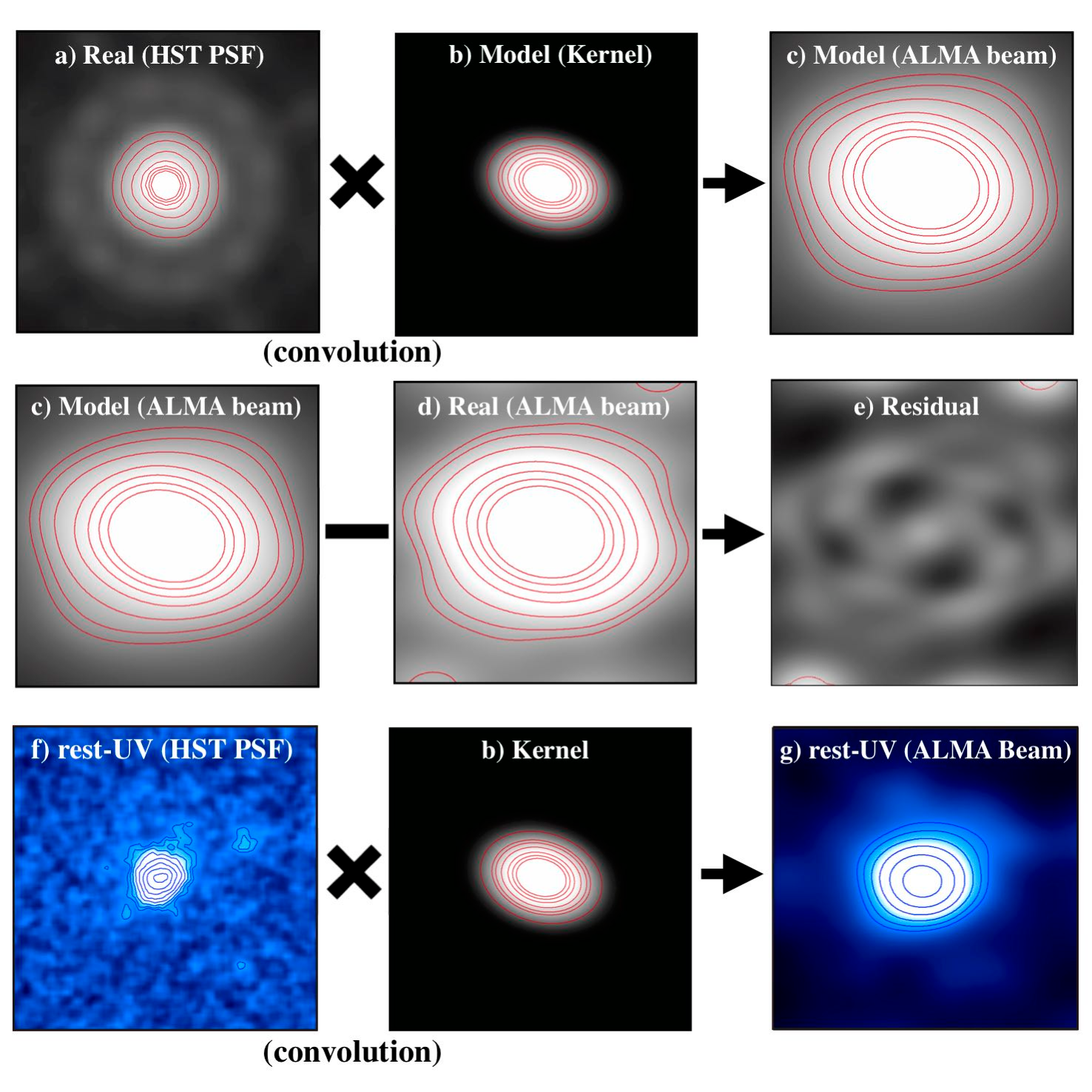}
 \caption[]{
Schematic overview to obtain the mock HST/H-band image whose spatial resolution is matched to the stacked ALMA image for the ALMA-HST sample: 
a) HST/H$-$band PSF,  
b) the best-fit kernel composed by three S$\acute{\rm e}$rsic profiles obtained with {\sc galfit},  
c) the best-fit ALMA beam model obtained with {\sc galfit}, 
d) the synthesized beam in the stacked ALMA image for the ALMA-HST sample, 
e) the residual between c) and d), 
f) the stacked HST/H$-$band image for the ALMA-HST sample, and 
g) the stacked HST/H$-$band image obtained by convolving f) with b). 
The red contours present 3\%, 5\%, 10\%, 20\%, 30\%, 40\%, and 50\% of the PSF or beam response. 
The blue contours denote the  2, 2$\sqrt{2}$, 4, ... $\times\sigma$ levels of the rest-frame UV continuum emission. 
The cutout sizes are $2''\times2''$ and $4''\times4''$ for the panels of a)$-$e) and f)$-$g), respectively. 
\label{fig:kernel}}
\end{center}
\end{figure*}

In panel (f) of Figure \ref{fig:kernel}, we show the stacked $H$-band image for the ALMA-HST sample. 
In the HST stacking, we adopted stacking centroids defined by the peak positions of the $H$-band images, 
smoothed with the $uv$-tapered ALMA beams in a consistent manner with the \cii\ line stacking.

To directly compare the size and morphology of the HST and ALMA images, 
we need to convolve the HST image to obtain a PSF that resembles the one of the ALMA image. 
We use {\sc galfit} to obtain a kernel with 
which the $H$-band PSF can be converted to the ALMA beam. 
For the kernel, we assume a sum of three independent S${\acute{\rm e}}$rsic profiles 
whose positions are fixed at the center. 

In Figure \ref{fig:kernel}, we present a schematic overview of converting the $H$-band PSF 
to the ALMA beam with the best-fit kernel. 
We first convolve the HST PSF (panel a) with the best-fit kernel (panel b) 
and derive the mock ALMA beam (panel c). 
We then subtract the actual ALMA beam (panel d) from the mock ALMA beam 
and produce the residual map (panel e). 
Within a radius of $1.''0$ on the residual map, we find that the difference between the mock and actual ALMA beams are less than $\sim 1.8$\%,  
showing that the best-fit kernel reproduces the ALMA beams well from the $H$-band PSF. 
We finally apply the convolution to the stacked $H$-band image (panel f) with the best-fit kernel, 
and obtain the mock $H$-band image whose PSF is almost the same as the stacked ALMA image.

\section{Results}
\label{sec:result}
\subsection{Discovery of {\sc [Cii]} Halo}
\label{sec:cii_halo}

\begin{figure*}
\begin{center}
\includegraphics[trim=0.4cm 0cm 0cm 0cm, clip, angle=0,width=0.8\textwidth]{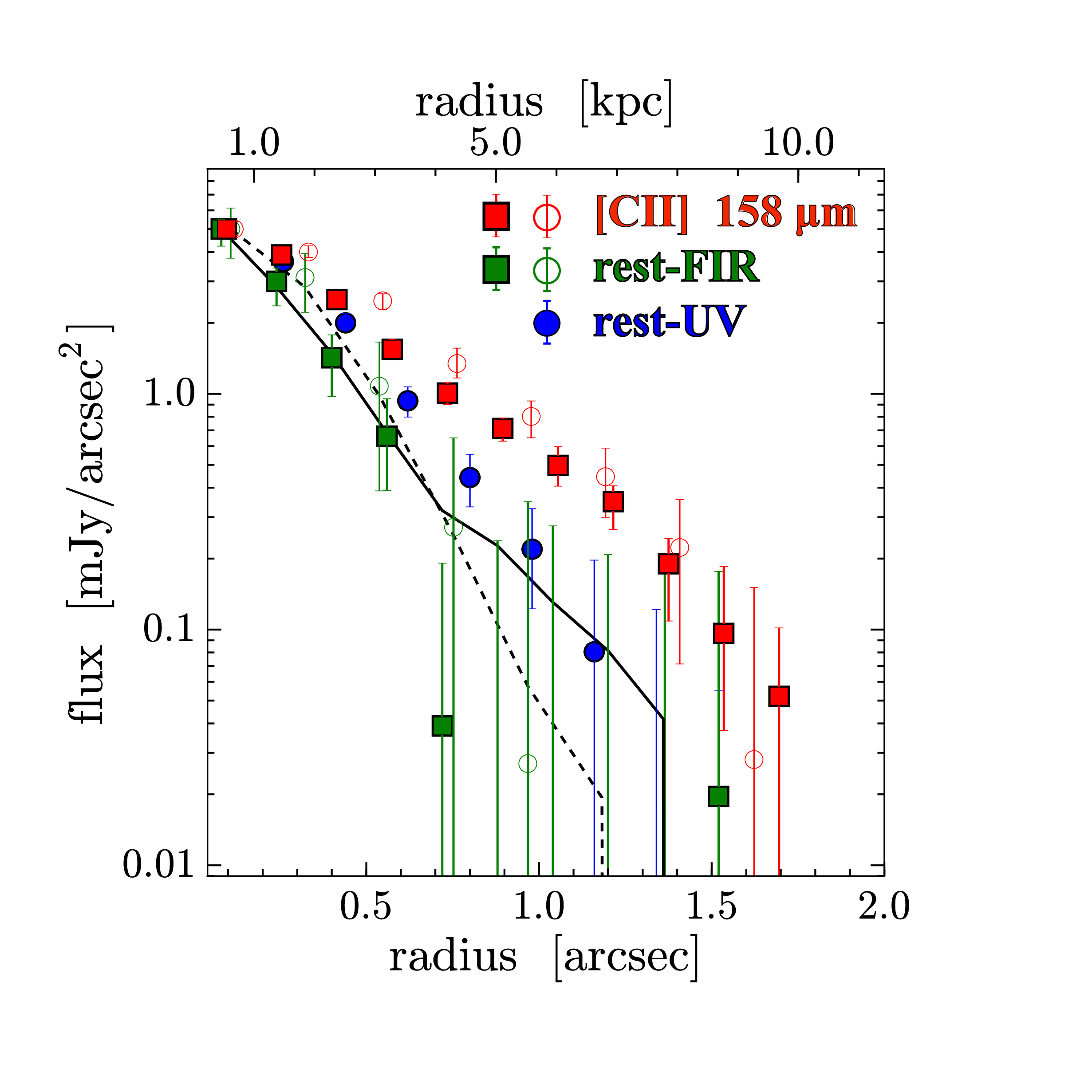}
 \vspace{-0.4cm}
 \caption[]{
Radial surface brightness profiles for the ALMA-HST (circles) and ALMA-ALL (squares) samples. 
The radial values are estimated by the median of each annulus. 
The red, green, and blue symbols denote the \cii\ line, rest-frame FIR, and rest-frame UV continuum emission. 
The rest-frame UV continuum profile is directly derived from the mock HST/$H$-band image whose resolution is matched to that of the ALMA image. 
The black dashed and solid curves denote the ALMA synthesized beams in the stacked images of the ALMA-HST and ALMA-ALL samples, respectively. 
All radial profiles are normalized to the peak value of the {\sc [Cii]} line. 
The green and red symbols are slightly shifted along the $x$-axis for clarity. 
\label{fig:radial_SB}}
\end{center}
\end{figure*}

Figure \ref{fig:radial_SB} presents the radial surface brightness profiles of the {\sc [Cii]} line, rest-frame FIR, and UV continuum, 
derived from the stacking results for the ALMA-HST (circles) and ALMA-ALL (squares) samples. 
For a fair comparison, the ALMA-HST results are obtained by re-performing the ALMA visibility-based stacking with the HST/$H$-band peak positions, 
while the ALMA-ALL results are not due to the lack of the HST/$H$-band images. 

In Figure \ref{fig:radial_SB}, 
the ALMA-ALL and ALMA-HST results show a good agreement in both profiles of the \cii\ line and the rest-FIR continuum. 
We find that the radial profile of the \cii\ line emission is extended up to a radius of $\sim$10 kpc 
which contrasts the rest-frame UV and FIR continuum. 
Because the typical effective radius of the normal star-forming galaxies at $z\sim6$ is estimated to be less than 1 kpc \citep[e.g.,][]{shibuya2015}, 
the $\sim$10-kpc scale structure at this epoch corresponds to the circum-galactic medium (CGM) surrounding the galaxies. 
These results suggest that the {\sc [Cii]} line emission is produced in the wide CGM areas even without stellar continuum. 
We discuss the physical origin of the {\sc [Cii]} halo in Section \ref{sec:discussion}. 

We also find that the profiles of the rest-frame FIR and UV continuum are consistent within the 1 $\sigma$ errors. 
Note that the rest-frame FIR continuum is likely to follow the ALMA beam, 
while the rest-frame UV continuum is slightly resolved with the ALMA beam. 
This suggests that the intrinsic size of the rest-frame FIR continuum is smaller than that of the rest-frame UV continuum, 
which is consistent with the recent ALMA results of the compact rest-frame FIR size more than the rest-frame UV and optical sizes 
among the star-forming galaxies at $z\sim2-4$ \citep[e.g.,][]{simpson2015a,ikarashi2015,hodge2016,fujimoto2017,fujimoto2018}. 

\subsection{Effect of \cii\--UV offset}
\label{sec:cii-uv_off}

Recent studies report a possibility that {\sc [Cii]}-line emitting regions 
are physically offset from the rest-frame UV ones \citep[e.g.,][]{maiolino2015}. 
To evaluate the potential effect from the {\sc [Cii]}-UV offsets in our results, 
we perform the ALMA and HST stacking for the ALMA-HST sample by
adopting two different stacking centers: 
HST/$H$-band and ALMA \cii\ line peak positions, and compare the radial profiles 
from these stacking results. 

\begin{figure}
\begin{center}
\includegraphics[angle=0,width=0.5\textwidth]{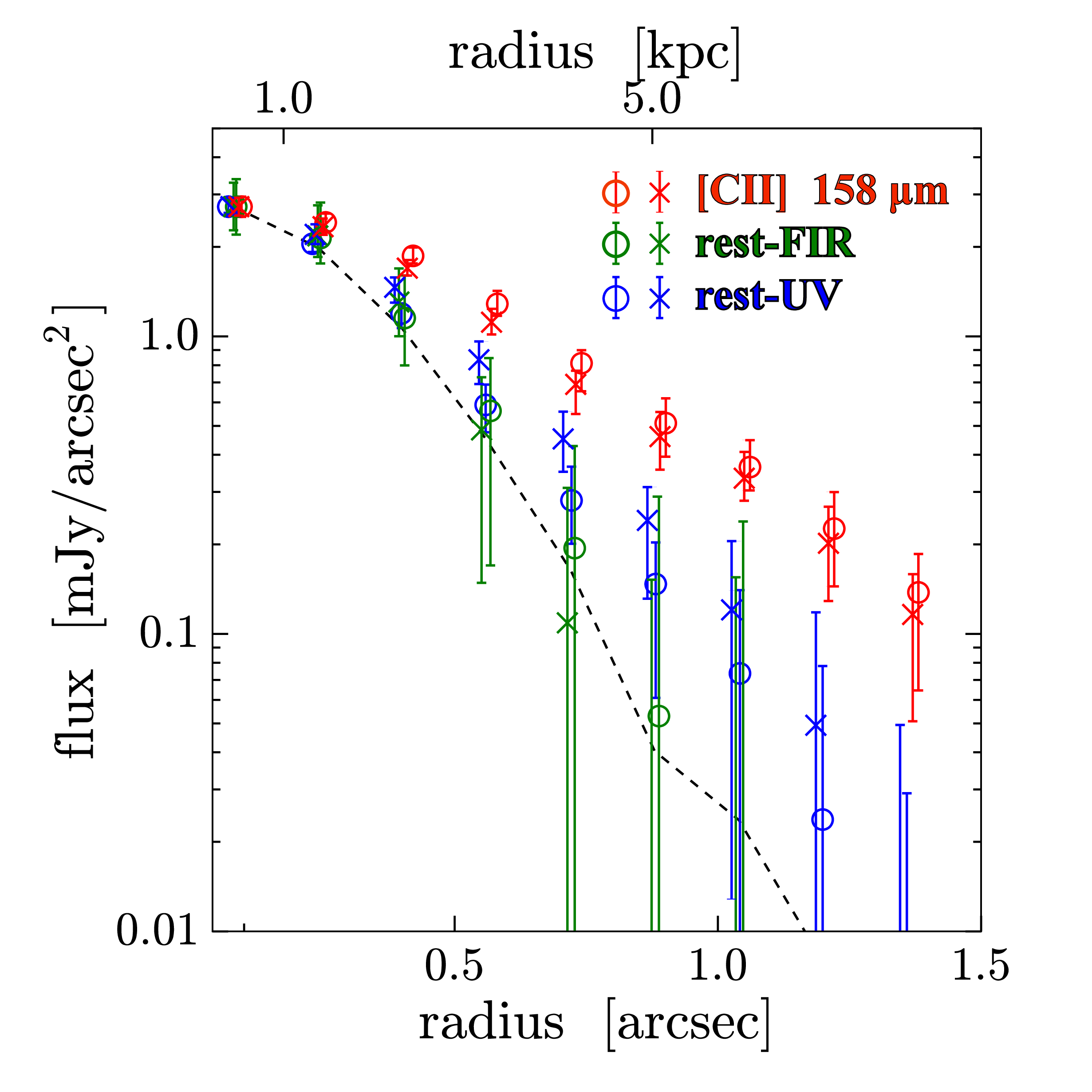}
\vspace{-0.4cm}
\caption[]{
Radial surface brightness profiles for the ALMA-HST sample derived with different stacking centers.
The red, green, and blue symbols denote the \cii\ line, rest-frame FIR, and rest-frame UV continuum emission. 
The color crosses and circles are the stacking results based on the stacking centers of the \cii\ line and the HST/$H$-band peak positions, respectively. 
The rest-frame UV continuum profile is directly derived from the mock HST/$H$-band image whose resolution is matched to the ALMA image. 
The black dashed curve denotes the ALMA synthesized beam. 
All radial profiles are normalized to the peak value of the \cii\ line.
The green and red symbols are slightly shifted along the $x$-axis for clarity. 
\label{fig:off_test}}
\end{center}
\end{figure}

In Figure \ref{fig:off_test}, the circle and cross symbols represent the stacking results derived with
the common stacking centers of the HST/$H$-band continuum and ALMA \cii\ line peak positions, respectively.  
We find that the \cii\ line profile is extended more than both the rest-frame FIR and UV continuum profiles in any cases. 
This suggests that the \cii\ line originates from much wider regions
than the continuum emission at rest-frame FIR and UV wavelengths, 
and clearly shows that the extended structure of the \cii\ line is not caused by the {\sc [Cii]}--UV offsets. 

\subsection{Radial ratio of $L_{\rm [CII]}$ to total SFR}
\label{sec:lcii}

To test whether the extended \cii\ line structure is caused by satellite galaxies, 
we investigate radial values of the \cii\ line luminosity $L_{\rm [CII]}$
at a given SFR derived from the rest-frame FIR and UV continuum. 
Because the ALMA-ALL and ALMA-HST results are consistent with each other (Figure \ref{fig:radial_SB}), 
we adopt the rest-frame UV results from the ALMA-HST sample, 
while the \cii\ line and rest-frame FIR continuum results from the ALMA-ALL sample 
to reduce the errors in the following estimates. 

\begin{figure*}
\begin{center}
\includegraphics[trim=0cm 1.0cm 0cm 1.0cm,clip,angle=0,width=1.0\textwidth]{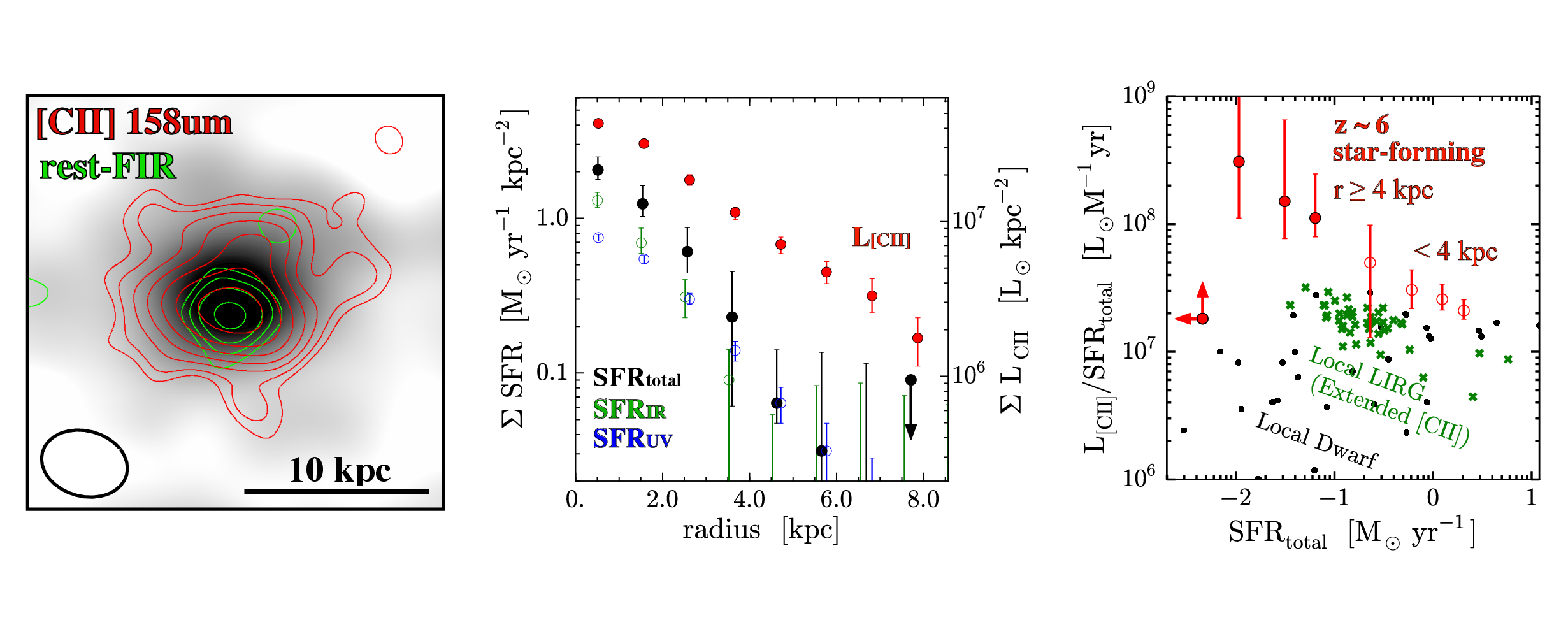}
\vspace{-0.4cm}
\end{center}
\caption{
{\it \bf Left:} 
Rest-frame UV emission of the ALMA-HST sample in the HST/$H$-band $4''\times4''$ image whose resolution is matched to the ALMA image. 
The red and green contours denote the 2, 2$\sqrt{2}$, 4, ... $\times\sigma$ levels of the {\sc [Cii]} line and the dust continuum emission, respectively. 
The ALMA synthesized beam is presented at the bottom left.
{\it \bf Middle:} 
Radial profiles of $\Sigma_{\rm SFR}$ (left axis) and $\Sigma_{L_{\rm [CII]}}$ (right axis). 
The blue, green, and black circles indicate $\Sigma_{\rm SFR_{UV}}$, $\Sigma_{\rm SFR_{IR}}$, and $\Sigma_{\rm SFR_{total}}$, respectively, 
based on Equations (1)--(3). 
The red circles denote $\Sigma_{L_{\rm [CII]}}$ normalized to $\Sigma_{\rm SFR}$ 
with the ratio of $L_{\rm [CII]}$/SFR$_{\rm total}$ = $10^{7}$ [$L_{\rm \odot}\,M_{\rm \odot}^{-1}\,{\rm yr}$] 
that is the average value in the local star-forming galaxies \citep{delooze2014}. 
{\it \bf Right:} 
Ratio of $L_{\rm [CII]}$/SFR$_{\rm total}$ as a function of SFR$_{\rm total}$. 
The filled (open) red circles indicate our stacking results at a radius of $\geq$ 4 kpc ($< 4$ kpc). 
The black dots denote local dwarf galaxy results in the global scale reported in \cite{delooze2014}.  
The green crosses present the extended \cii\ line emission calculated from the local LIRG results in \citep{diaz-santos2014}. 
We assume the area of 1 kpc$^{2}$ for the $L_{\rm [CII]}$ and SFR$_{\rm total}$ estimates in our stacking and the local LIRGs results.
At the radius of $>7$ kpc, 
the SFR$_{\rm total}$ value in our stacking results becomes negative due to the noise fluctuations on the low surface brightness of the rest-frame UV and FIR continuum emission, 
where we evaluate the lower limit of the ratio by using the upper limit of SFR$_{\rm total}$.
\label{fig:radial_ratio}}
\end{figure*}

We first estimate the radial $L_{\rm [CII]}$ value.   
For our sources, 
the weighted-average source redshift and FWHM of the \cii\ line width are estimated to be $z=6.01$ and 270 km s$^{-1}$, respectively. 
Since the velocity-integrated width is 100 km s$^{-1}$ in the stacked \cii\ line map, 
we correct the velocity-integrated value in the range from 100 km s$^{-1}$ to 270 km s$^{-1}$, 
assuming a single Gaussian line profile, to recover the total value of $L_{\rm [CII]}$. 
We second evaluate the radial SFR value. 
We derive the obscured (SFR$_{\rm IR}$), un-obscured (SFR$_{\rm UV}$), 
and total SFR (SFR$_{\rm total}$) with the equations in \cite{murphy2011} of 
\begin{eqnarray}
{\rm SFR_{IR}} \, [M_{\odot}\,{\rm yr^{-1}}] &=& 3.88\times10^{-10} \, L_{\rm IR} \, [{\rm erg\,s^{-1}}], \\
{\rm SFR_{UV}} \, [M_{\odot}\,{\rm yr^{-1}}] &=& 4.42\times10^{-10} \, L_{\rm UV} \, [{\rm erg\,s^{-1}}], \\
{\rm SFR_{total}} &=& {\rm SFR_{IR}} + {\rm SFR_{UV}}, 
\end{eqnarray}
where $L_{\rm IR}$ is the integrated IR flux density estimated by a typical modified blackbody 
whose spectral index $\beta_{\rm d}$ and dust temperature $T_{\rm d}$ are 
$\beta_{\rm d}=1.8$ \citep{planck2011} and $T_{\rm d}=35$ K \citep{coppin2008}, 
and $L_{\rm UV}$ is the rest-frame UV luminosity at 0.16 $\mu$m with the HST $H$-band. 
Finally, we divide the radial $L_{\rm [CII]}$ values 
by the radial SFR values 
and obtain the radial ratio of $L_{\rm [CII]}$/SFR$_{\rm total}$. 

In Figure \ref{fig:radial_ratio}, 
we show the surface densities of $L_{\rm [CII]}$ ($\Sigma_{L_{\rm [CII]}}$) and SFR$_{\rm total}$  ($\Sigma_{\rm SFR}$) as a function of radius (middle panel), 
and the radial ratio of $L_{\rm [CII]}$/SFR$_{\rm total}$ (right panel) as a function of SFR$_{\rm total}$ for our stacking results. 
For comparison, the right panel of Figure \ref{fig:radial_ratio} also presents global scale $L_{\rm [CII]}$/SFR$_{\rm total}$ ratios of the local dwarf galaxies \citep{delooze2014}. 

In the right panel of Figure \ref{fig:radial_ratio}, 
the red filled and open circles denote our stacking results in the outer (radius of $\geq$ 4 kpc) and central ($<$ 4 kpc) regions, respectively. 
We find that $L_{\rm [CII]}$/SFR$_{\rm total}$ decreases with SFR$_{\rm total}$. 
The highest ratios ($> 10^{8}\, L_{\odot}\,M_{\odot}^{-1}\,{\rm yr}$) are found at the outer regions 
and not compatible with typical values found in the local dwarf galaxies ($<3\times10^{7}\, L_{\odot}\,M_{\odot}^{-1}\,{\rm yr}$; black dots in the figure). 
These results indicate that the \cii\ halo is not likely driven by satellite galaxies. 
Note that other types of high-$z$ galaxies with SFR$_{\rm total}$ $>$ 10 $M_{\odot}$ such as star-forming, submillimeter, and quasar-host galaxies show 
ratios of 10$^{6}\sim10^{7}$ $L_{\odot}\,M_{\odot}^{-1}\,{\rm yr}$ \citep[e.g.,][]{capak2015,rybak2019,venemans2019}  
which are yet difficult to explain the highest ratios in our stacking results ($> 10^{8}\, L_{\odot}\,M_{\odot}^{-1}\,{\rm yr}$). 

\cite{diaz-santos2014} report the \cii\ line emission extended over $\sim1-10$ kpc scale around local luminous infrared galaxies (LIRGs), 
and we also compare our results with this spatially resolved data. 
In the right panel of Figure \ref{fig:radial_ratio}, we show the $L_{\rm [CII]}$/SFR$_{\rm total}$ ratios of the extended emission around local LIRGs. 
We find that the highest ratios in our stacking results ($> 10^{8}\, L_{\odot}\,M_{\odot}^{-1}\,{\rm yr}$) are still higher than those around the local LIRGs ($<3\times10^{7}\, L_{\odot}\,M_{\odot}^{-1}\,{\rm yr}$; green crosses in the figure).  
The \cii\ halo at $z\sim6$ is thus not identical to the extended \cii\ line emission observed in the local Universe, 
which may suggest that \cii\ halos evolve with redshift.
We discuss possible origins of the \cii\ halo in Section \ref{sec:discussion}. 

\subsection{Scale Length of {\sc [Cii]} Halo}
\label{sec:halo_size}
We characterize the detail radial surface brightness profile of the {\sc [Cii]} line emission by two-component fitting with {\sc galfit}. 
Here we assume the two components as the central and the halo components. 

For the central component, 
we adopt the S$\acute{\rm e}$rsic profile whose parameters are estimated from the rest-frame UV profile in the stacked HST/$H$-band image (Figure \ref{fig:kernel} f). 
We obtain the best-fit effective radius $r_{\rm e}$ and the S$\acute{\rm e}$rsic index $n$ of $r_{\rm e}=1.1\pm0.1$ kpc and $n=1.2\pm0.01$ that are consistent with the average values estimated from the normal star-forming galaxies at $z\sim6$ \citep{shibuya2015}. 
For the halo component, we utilize the exponential profile. 
The exponential profile has been used for scale-length measurements of the Ly$\alpha$ halo which 
is universally identified around the high-z star-forming galaxies \citep[e.g.,][]{steidel2011,matsuda2012,momose2014,momose2016,leclercq2017}. 
The exponential profile is described as $C_{\rm n}$exp($-r/r_{\rm n}$) where $C_{\rm n}$ is a constant and $r_{\rm n}$ is the scale length. 
We fix the central positions of both central and halo components to obtain a stable result. 

\begin{figure}[h]
\begin{center}
\includegraphics[trim=0cm 0cm 0cm 0cm, clip, angle=0,width=0.5\textwidth]{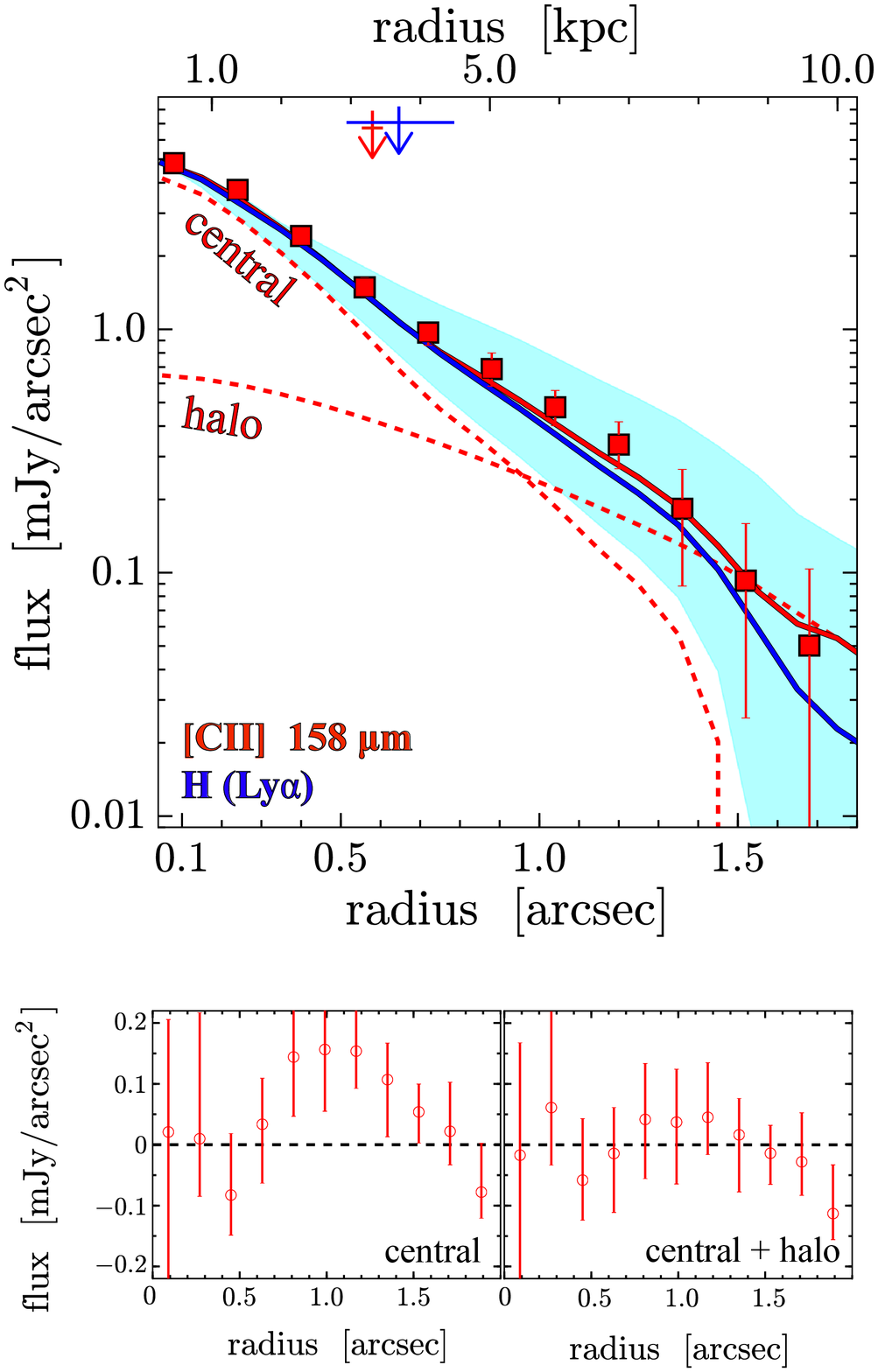}
 \vspace{-0.2cm}
 \caption[]{
{\bf Top:} 
Two-component S$\acute{\rm e}$rsic+exponential profile fitting for the \cii\ line, averaged over 18 galaxies.
The red-dashed curves represent the best-fit results of the central stellar continuum and outer halo components, 
while the solid red curve denotes the sum of the best-fit two-component results. 
The solid blue curve and the shaded region indicate the median and the 16--84th percentile of the
radial surface brightness profile of the Ly$\alpha$ lines in a recent control sample from MUSE \citep{leclercq2017}. 
For the Ly$\alpha$ line, we convolve the best-fit results of the S$\acute{\rm e}$rsic+exponential profiles with the ALMA beam. 
The red and blue arrows with error bars show the best-fit scale lengths of the \cii\ and the Ly$\alpha$ halo components, respectively. 
{\bf Bottom:} 
Residuals in the best-fit results of one- (left) and two-component (right) profile fittings. 
\label{fig:2comp}}
\end{center}
\end{figure}

Top panel of Figure \ref{fig:2comp} presents the best-fit results with the S$\acute{\rm e}$rsic+exponential profiles for the {\sc [Cii]} line emission.
We obtain the best-fit scale-length values of $r_{\rm n}=$ 3.3 $\pm$ 0.1 kpc. 
This corresponds to the best-fit effective radius of $r_{\rm e}=5.6$ $\pm$ 0.1 kpc, 
showing that the \cii\ halo is extended $\sim$5 times more than the stellar continuum in the central galactic component.  
{
Note that the visibility-based profile fitting with {\sc uvmultifit} \citep{marti2014} also provides us with the best-fit value of $r_{\rm e}=5.1\pm$ 1.7 kpc for the halo component, 
which is consistent with the {\sc galfit} result within the error. 
}

In Figure \ref{fig:2comp}, 
we compare the radial surface brightness profiles of the {\sc [Cii]} with the Ly$\alpha$ halos universally identified in the normal star-forming galaxies at $z\sim3$--6 \citep[e.g.,][]{momose2016,leclercq2017}. 
For the {\sc [Cii]} line emission, we adopt the result from the ALMA-ALL sample due to the high significance detection. 
For the Ly$\alpha$ line emission, we use the recent results with the deep MUSE data for the high-$z$ LAEs of \cite{leclercq2017}, 
where the authors estimate the best-fit radial surface brightness profiles by fitting the two-component S$\acute{\rm e}$rsic+exponential profile. 
We select the best-fit results of 6 LAEs at $z>5$ with $M_{\rm UV}\lesssim-21$ mag and EW$_{\rm Ly\alpha}$ $<$ 100 ${\rm \AA}$ that are consistent with the parameter space of our sample (Table \ref{tab:our_alma}). 
We find that the radial surface brightness profile of the {\sc [Cii]} line emission is comparable to that of the Ly$\alpha$ line emission. 
The median $r_{\rm n}$ value for the 6 LAEs is estimated to be $3.8\pm1.7$ kpc that is consistent with our best estimate of $3.3\pm0.1$ kpc. 
These results may imply that the physical origin of the extended {\sc [Cii]} line emission is related to the Ly$\alpha$ halo. 

Note that we confirm that it is hard to reproduce the extended morphology of the \cii\ line emission with the central component alone.   
In the bottom panel of Figure \ref{fig:2comp}, we present the residuals of the \cii\ line emission obtained from the best-fit results of
the one- (central) and two- (central+halo) component fittings with {\sc uvmultifit}. 
We find that the residuals in the one-component fitting result show a bump at a radius of $\sim1''$ over the errors, 
while the residuals in the two-component fitting result is broadly consistent with zero. 
This suggests that the extended morphology of the \cii\ line emission 
consists of a combination of the central plus halo components.

\subsection{\cii\ Stacked Spectrum}
\label{sec:stack_spectrum}

We also perform the stacking for the \cii-line spectra of the ALMA-ALL sample 
to test whether our ALMA sample has a broad wing feature 
which is a good probe for the on-going outflow activities. 
For the stacking procedure, we adopt the same manner as previous ALMA studies \citep{decarli2018,bischetti2018}. 
Here we adopt a relatively small aperture diameter of $0\farcs4$ for the individual spectra to reduce the contamination of the close companions \citep{jones2017,carniani2018b}.

\begin{figure}
\begin{center}
\includegraphics[trim=0.2cm 0cm 0.2cm 0cm, clip,angle=-90,width=0.5\textwidth]{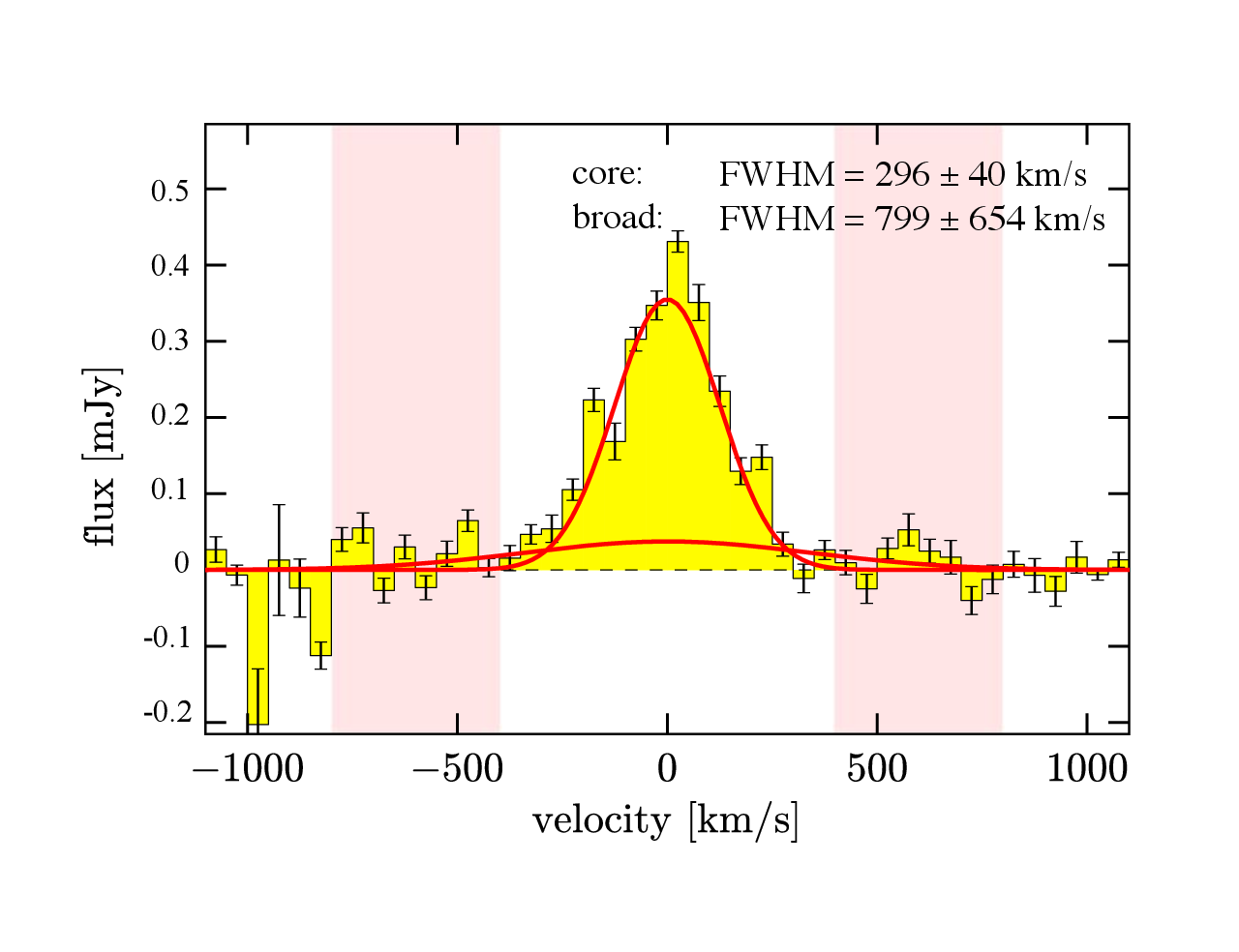}
\end{center}
\vspace{-0.4cm}
\caption{
ALMA \cii-line spectrum averaged over the ALMA-ALL sample. 
The spectrum is derived with an aperture diameter of $0\farcs4$. 
The red curves denote the best-fit two (= core + broad) Gaussian component model. 
The shade regions indicate the velocity ranges in which the velocity-integrated intensity is tentatively detected 
at the 3.2$\sigma$ level.  
\label{fig:stack_spectrum}}
\end{figure}

In Figure \ref{fig:stack_spectrum}, we show the stacked \cii-line spectrum with the best-fit two Gaussian component model:
the combination of the core and broad components whose velocity centers are fixed at 0 km/s for the stable results.  
The best-fit FWHMs are estimated to be 296 $\pm$ 40 km/s and 799 $\pm$ 654 km/s for the core and broad components, respectively. 
In the velocity range of $\pm$ 400 -- 800 km/s, 
the velocity-integrated intensity is tentatively detected at the 3.2$\sigma$ level. 
Moreover, \cite{sugahara2019} have recently reported that the rest-frame UV metal absorption lines are blue-shifted with the central outflow velocity of 440$^{+110}_{-140}$ km s$^{-1}$ 
from the \cii-systemic redshift in the stacked Keck spectra whose stacking sample includes 6 out of our 18 \cii\ line sources. 
These results may suggest the existence of the tentative broad wing feature is produced by the outflow. 

Note that there are other possibilities that produce the broad wing feature.
One possibility is that the contamination of the satellite galaxies. 
The \cii\ line emission from the individual satellite galaxies can be smoothed in the stacking procedure for the 18 galaxies, 
which may be identified as the broad wing feature. 
Another possibility is that the faint continuum emission is mistakenly identified as the broad wing feature. 
Although we have performed the continuum subtraction for the \cii\ line data cubes of the 4 galaxies whose continuum emission is individually detected, 
it is possible that the faint continuum emission from the rest of the 14 ($=$ 18 $-$ 4) galaxies appears in the deeply stacked spectrum. 
Since the significance of the broad wing is low, 
we cannot draw definite conclusions from our data. 

\subsection{Comparison with Model}

We compare our observational results with two independent numerical simulations for star-forming galaxies 
with the halo mass of $M_{\rm halo}\sim 10^{11}-10^{12}$\,M$_{\odot}$ at $z\sim6$.
Note that our sample is characterized by the average $M_{\rm UV}$ value of $\lesssim -21$ mag (Table 1) 
which corresponds to $M_{\rm halo}\approx 10^{11}-10^{12}M_{\odot}$ from the $M_{\rm UV}$--$M_{\rm halo}$ relation \citep{harikane2018}. 

First set is a zoom-in simulation for a star-forming galaxy, 
Alth$\ae$a \citep{pallottini2017b,pallottini2017a,behrens2018}.
The hydrodynamical and dust radiative transfer (RT) simulations are combined, 
which provides realistic predictions for the spatial distribution of the \cii\ line as well as the rest-frame 
FIR and UV continuum emission, with a spatial resolution of 30\,pc. 
The hydrodynamical and the dust RT simulations are fully described in previous studies \citep{pallottini2017b,pallottini2017a,behrens2018}. 
Note that the dust RT is calculated as a post-processing step on snapshots of the hydrodynamical simulation. 
The \cii\ line emission is computed in post-processing \citep{vallini2015} by adopting the photoionization code, 
{\sc cloudy} \citep{ferland2017}. In these processes, CMB suppression \citep{dacunha2013,zhang2016,pallottini2015,lagache2018} is included in the calculation.

{
Second set is another cosmological hydrodynamic zoom-in simulations performed by the smoothed particle hydrodynamics (SPH) code {\sc Gadget-3} \citep{springel2005a} with the sub-grid models developed in 
{\it Overwhelmingly Large Simulations} (OWLS) project \citep{schaye2010} and the {\it First Billion Year} (FiBY) project \citep[e.g.,][]{johnson2013} which reproduce the general properties of the high-redshift galaxy population well 
\citep[e.g.,][]{cullen2017}. 
For this comparison, we use four different halos: Halo-12, Halo-A, Halo-B, and Halo-C that have $M_{\rm halo} = (0.5-1) \times 10^{12} M_{\odot}$ at $z=6.2-6.5$. 
The details of Halo-12 is discussed in 
\citep{yajima2017,arata2019b,arata2019}, but the latter three halos are newly simulated for this paper with similar initial conditions but with different merger histories. 
The minimum gravitational softening length is $\epsilon_g = 200$\,pc (comoving), therefore we achieve $\sim$25\,pc resolution at $z=7$ for gravity.  We also allow the SPH smoothing length to adaptive down to 10\% of $\epsilon_g$, therefore the hydrodynamic resolution reaches a several parsecs at $z=6-7$.    
The RT calculation including the dust absorption/re-emission is performed as a post-process with ``All-wavelength Radiative Transfer with Adaptive Refinement Tree" (ART$^{2}$ code: \citealt{li2008,yajima2012}).
This calculation provides the SED over a wide wavelength range, and solves for the ionization structure of ISM/CGM.  The \cii\ line emissivity is estimated from the ionized carbon abundance. 
The details of hydrodynamic simulations, RT, and the \cii\ line computations are fully described in \cite{yajima2017}, \cite{arata2019}, and Arata et al. (2019, in prep.).  
}

\begin{figure*}
\begin{center}
\includegraphics[trim=0cm 0cm 0cm 0cm, clip, angle=90,width=1.\textwidth]{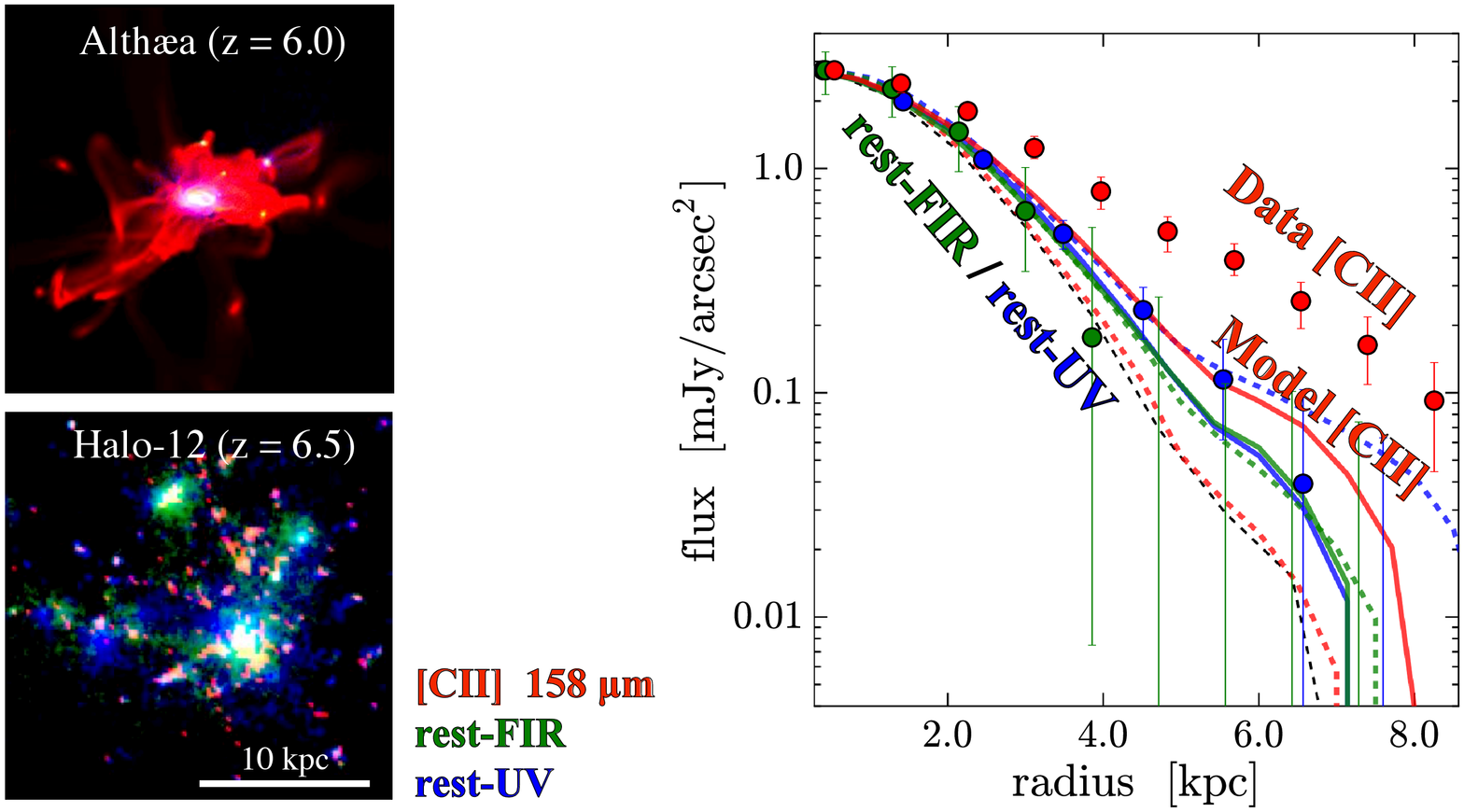}
\vspace{-1.6cm}
\caption[]{
{
{\it \bf Left:} 
$4''\times4''$ fake-color image for Alth$\ae$a at $z=6.0$ (top) and Halo-12 (bottom) in the zoom-in simulations 
(red: {\sc [Cii]} line, green: rest-frame FIR continuum, blue: rest-frame UV continuum). 
{\it \bf Right:} 
Radial surface brightness profiles of the {\sc [Cii]} line (red curve), rest-frame FIR (green curve), and UV (blue curve) continuum emission estimated in the zoom-in simulations via stacking procedure. 
The solid and dashed color lines present the Alth$\ae$a and Halo-12 results, respectively.  
The black dashed curve denotes the ALMA synthesized beam. 
The circles indicate the ALMA-HST stacking results whose colors are assigned in the same manner as the left panel.  
} 
\label{fig:model}}
\end{center}
\end{figure*}

The left panel of Figure \ref{fig:model} presents a color composite of the {\sc [Cii]} line, rest-frame UV, and FIR continuum emission of Alth$\ae$a at $z=6.0$ 
and Halo-12 at $z=6.5$ in the zoom-in simulations.
The surface brightness morphology of the {\sc [Cii]} line emission clearly shows the extended structure over the 10-kpc scale with surrounding satellite clumps and filamentary structures. 
The picture of the extended \cii\ halo around the central galaxies is roughly consistent with the observational results. 

To quantitatively compare the zoom-in simulation to our observational results, 
we carry out the stacking for the zoom-in simulation results of the \cii\ line, 
the rest-frame FIR and the rest-frame UV continuum in the same manner as the observations.
For the Alth$\ae$a simulation, 
we take 12 snapshots at different redshifts within $6.0\leq z\leq7.2$  
For each snapshot, we calculate the surface brightness from face-on and three random angles, 
where the \cii\ line emissivity is calculated within 100\,km\,s$^{-1}$ of the velocity center of the galaxy to match the visibility-based stacking procedure for the ALMA data.  
In this way, we obtain 48 (=$12\times4$) images of Alth$\ae$a. 
We refer to \cite{kohandel2019} for a full analysis of the different morphological results from different evolutionary stages and viewing angles. 
We then select 9 out of 48 snapshots randomly -- the same sample size used for the stacking of the ALMA-HST sample. 
Finally, we perform the stacking of the intrinsic images, and smooth the stacked images with the ALMA beam.
For the second set of simulation, we calculate the surface brightness from three orthogonal angles 
for four different halos (Halo-12, Halo-A, Halo-B, Halo-C) and obtain 12 (=$4\times3$) images.  
We then carry out the same stacking and smoothing procedures as the first simulation. 

In the right panel of Figure \ref{fig:model}, 
we show the radial surface brightness profiles estimated from the two independent zoom-in simulations.  
For comparison, we also plot the observational results obtained in Section \ref{sec:result}. 
We find that both simulations reproduce the overall trend of observational results of rest-frame UV and FIR continuum within the errors. 
However, we also find that the {\sc [Cii]} line emission in both simulations is not as extended as the observed data. 
In Alth$\ae$a, although it reproduces the trend that the \cii\ line is more extended than that of the rest-frame FIR and UV continuum,  the intensity of the \cii\ emission at $r > 5$ kpc is still lower than the observed one. 
In Halo-12, the \cii\ line is the least extended. 
These results indicate that the existence of the \cii\ halo challenges current hydrodynamic simulations of galaxy formation. 

\section{Discussion}
\label{sec:discussion} 
In Section \ref{sec:result}, 
we find that the {\sc [Cii]} line emission is extended up to $\sim$10-kpc scale around the normal star-forming galaxies at $z=5-7$ 
and is potentially related to the Ly$\alpha$ halo. 
In contrast to the previous reports of the 10-kpc-scale carbon reservoirs around rare, massive galaxies, such as dusty starbursts and quasars at $z\sim2$--6 \citep[e.g.,][]{ivison2011,falgarone2017,george2014, diaz-santos2014,maiolino2012,cicone2015}, 
our results indicate that the cold carbon gas halo universally exists even around early normal galaxies. 

\begin{figure*}
\begin{center}
\includegraphics[trim=0cm 0cm 0cm 0cm, clip, angle=0,width=1.\textwidth]{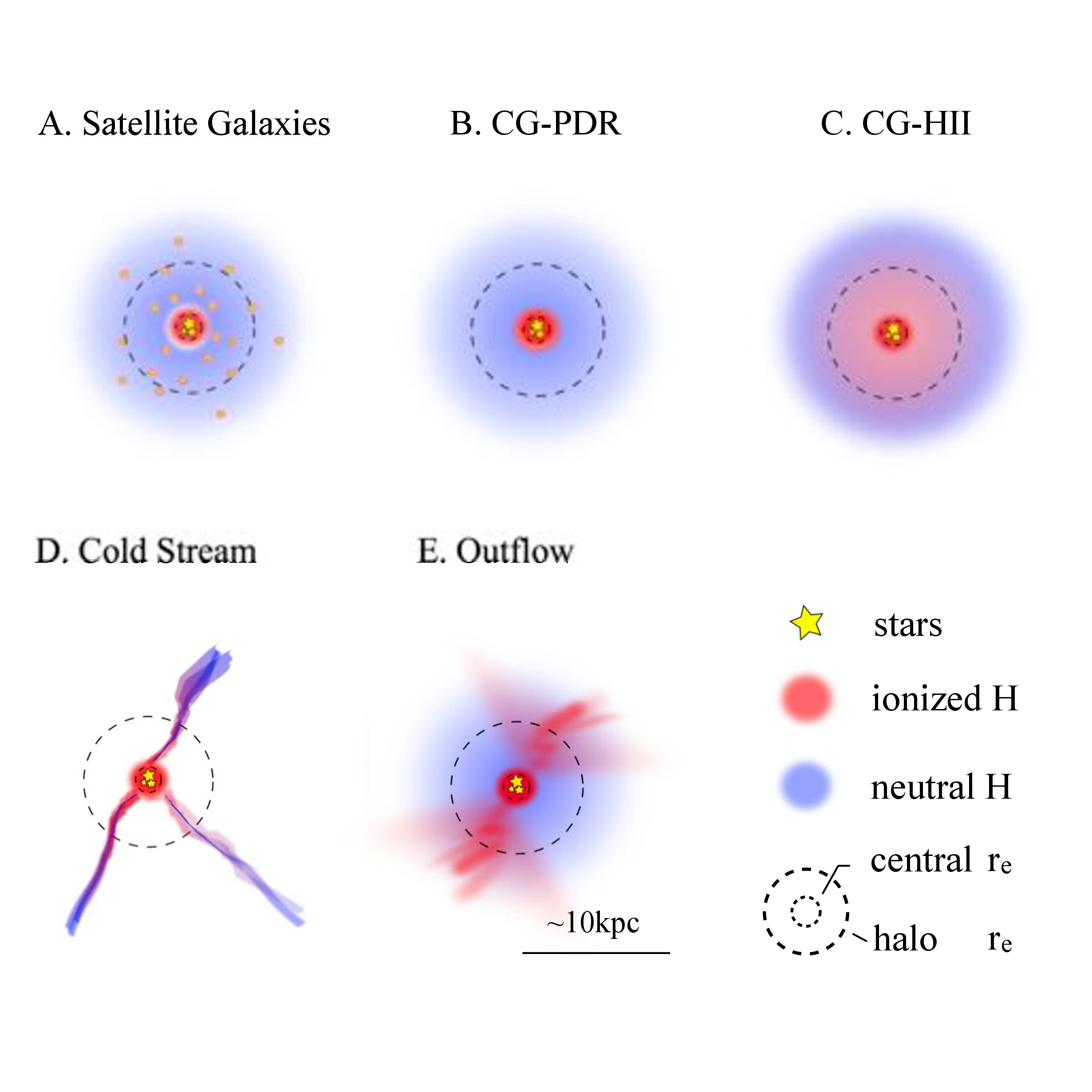}
 \vspace{-0.6cm}
 \caption[]{
Illustrations of five possible scenarios for the physical origin of the \cii\ halo with the potential association of the Ly$\alpha$ halo:
A) satellite galaxies; B) circum-galactic (CG) photodissociation region (PDR); C) {CG HII region (CG-HII)}; D) cold stream; and E) outflow. 
The blue and red shades show the neutral and ionized hydrogen in ISM and CGM. 
The yellow stars represent the star-forming regions. 
The inner and outer dashed circle denote the effective radii ($r_{\rm e}$) of the central and halo components of the {\sc [Cii]} line emission, respectively.
\label{fig:scenario}}
\end{center}
\end{figure*}

The existence of the cold carbon gas halos around the early normal galaxies
raises two questions: what powers the \cii\ line emission 
and how is the carbon abundance in the circum-galactic (CG) area
enriched at such early cosmic epochs.  
Theoretical studies suggest the following five scenarios 
that can give rise to the extended \cii\ line emission
with the potential association of the Ly$\alpha$ halo:  \\
\indent
A) satellite galaxies,  \\
\indent
B) CG photodissociation region (PDR), \\
\indent
C) CG HII region, \\ 
\indent
D) cold streams, \\
\indent
E) outflow.\\
These five scenarios are illustrated in Figure \ref{fig:scenario}. 

The first scenario invokes satellite galaxies (Figure \ref{fig:scenario}-A). 
If satellite galaxies exist around the central star-forming galaxies, 
the {\sc [Cii]} and Ly$\alpha$ line emission from the satellite galaxies will be observed as extended structures around the central galaxies. 
In this scenario, the extended halo size is determined by the spatial distribution of the satellite galaxies, 
which explains both extended components of the {\sc [Cii]}  and Ly$\alpha$ line emission. 

The second scenario is a PDR extended over CG scale, referred to as CG-PDR (Figure \ref{fig:scenario}-B). 
The ionizing photons ($h\nu>13.6$ eV) from massive stars form the HII region on the central galactic scale. 
Far-ultraviolet (FUV) photons (6 eV $<h\nu< 13.6$ eV) penetrate the surrounding ISM deeper than the ionizing photons, 
making the PDR more extended than the HII region. 
In these PDRs, the carbon is still singly ionized (11.3 eV) by the FUV photons.  
If the PDR is extended over the CG scale, the extended {\sc [Cii]} line emission is thus detected on the CG scale. 
Besides, the Ly$\alpha$ line emission is also spatially extended due to the resonant scattering by the neutral hydrogen in the surrounding ISM \citep[e.g.,][]{xue2017}. 

The third scenario is that ionizing photons penetrate the surrounding ISM deeper and form large HII regions even spreading over the CGM, which we refer to as CG-HII 
(Figure \ref{fig:scenario}-C). 
This scenario is similar to scenario (B), 
but the existence of strong ionizing sources and/or ISM properties differ from scenario (B), 
and the HII region is larger than scenario (B) where the carbon is singly ionized. 
In this case, 
the Ly$\alpha$ line emission is extended due to the fluorescence \citep[e.g.,][]{mas-ribas2016}, instead of the resonance scattering in scenario (B).

The fourth scenario is cold streams (Figure \ref{fig:scenario}-D). 
Cosmological hydrodynamical simulations suggest that intense star-formation in high-$z$ galaxies is fed by a dense and cold gas ($\sim$10$^{4}$ K) 
which is dubbed cold streams \citep[e.g.,][]{dekel2009}. 
The cold streams radiate {\sc [Cii]} as well as Ly$\alpha$ line emission powered by gravitational energy, 
and produce the extended {\sc [Cii]} and Ly$\alpha$ line emission around a galaxy. 
Moreover, the cold stream may cause shock heating which can also produce the {\sc [Cii]} and Ly$\alpha$ line emission. 

The fifth scenario is outflow (Figure \ref{fig:scenario}-E). 
In the outflow, the ionized carbon and hydrogen powered by the AGN and/or star-formation feedback produce the extended {\sc [Cii]} and Ly$\alpha$ line emission (see also \citealt{faisst2017}). 
The associated process of the shock heating may also contribute to radiating these line emission. 
Note that although we choose ALMA sources not reported as AGNs, 
we cannot rule out the possibility that our ALMA sources contain faint AGNs and/or have the past AGN activity. 

In the following subsections, 
we discuss these possibilities based on the observational and theoretical results. 

\subsection{Hints From Observational Results}
\label{sec:obs}

In the observational results, 
the \cii\ line is more extended than both the rest-frame FIR dust and UV continuum
beyond the errors up to a radius of at least $\sim7$ kpc (Figure \ref{fig:radial_SB}). 
Assuming a constant \cii\ line emissivity at a given stellar continuum \citep{delooze2014}, 
the large gap between the radial profiles of the \cii\ line and the stellar continuum indicates that 
the stellar continuum is not enough to explain the large part of the \cii\ line emissivity of the \cii\ halo. 

Although the \cii\ line emissivity may be changed from the central to halo areas at a given stellar continuum, 
the metallicity at such outer areas is expected to be $\sim1\%$ of the galaxy center \citep{pallottini2017b}. 
Even if the stellar continuum in the outskirts is coming  from low-mass, faint satellite galaxies, 
the mass$-$metallicity relation \citep{mannucci2010} also  suggests lower metallicities for the satellite galaxies. 
Because lower metallicity reduces the \cii\ line emissivity for a given stellar continuum \citep{vallini2015},
it would be difficult to explain the \cii\ halo by the same source as the stellar continuum. 
In fact, Figure \ref{fig:radial_ratio} shows that the $L_{\rm [CII]}$/SFR$_{\rm total}$ ratio becomes higher towards outskirts of the halo, which cannot be explained by the dwarf galaxies. 
Our observational results thus rule out scenario (A), and support the other four scenarios. 

In the recent {\sc [Cii]} line studies at $z>5$, 
\cite{gallerani2018} report signatures of starburst-driven outflows from 9 normal star-forming galaxies at $z\sim5.5$ with the stacked {\sc [Cii]} spectra. 
With the similar sample, the rest-frame UV metal absorptions are also identified to be blue-shifted from the \cii-systemic redshift in the stacked Keck spectra \citep{sugahara2019}.  
From more luminous objects, 
the broad wing features are detected in the stacked \cii\ line spectra of $z\sim4-6$ quasars \citep{bischetti2018} as well as in an individual \cii\ line spectrum of a quasar at $z=6.4$ \citep{maiolino2012, cicone2015}. 
These recent results suggest that scenario (E) may be potentially very interesting.  
However, it is also reported that the stacked {\sc [Cii]} spectra even from 23 quasars at $z\gtrsim6$ shows no clear evidence of the existence of strong feedback \citep{decarli2018}. 
Also, our stacked \cii\ spectra with the 18 star-forming galaxies does not show a clear broad wing feature, neither (Section \ref{sec:stack_spectrum}). 
Among scenarios of (B), (C), (D), and (E), we thus cannot conclude the most likely one from our and recent observational results. 

\subsection{Hints From Theoretical Results}
\label{sec:model}

In the simulation results, 
the extended profile of \cii\ halo is not fully reproduced (Figure \ref{fig:model}). 
{
This may suggest that some physical processes are not sufficiently solved in the simulations, e.g., metal enrichment, feedback, ISM/CGM clumpiness, and the propagation of ionizing radiation.
On the other hand,}
if the current assumptions related to the \cii\ line emissivity are correct, 
additional mechanism(s) are required to produce the extended \cii\ line emissivity in the simulation. 

There are two possibilities for such additional mechanisms that are not included in the calculation of the \cii\ line emissivity in the simulations. 
The first possible mechanism is the shock heating;  {\citet{appleton2013} have shown that \cii\ can be excited on large scales from the dissipation of mechanical energy of galaxy mergers via turbulent cascade. 
Although the shock heating should be captured in the hydrodynamical calculation of the zoom-in simulation, it is possible that the current simulations do not have sufficient resolution to capture the turbulent cascade of large-scale mechanical energy down to the molecular cloud scales. 
This means that the computation of the ionized carbon abundance in {\sc cloudy} does not adequately consider the effect of shock heating. }
Since shock heating is caused by galaxy merger or gas inflow/outflow processes, 
the \cii\ emissivity could become more enhanced if the shock heating and associated turbulent cascade is properly treated in the scenarios (A), (D), and (E).

The second possible mechanism is the past/on-going AGN activities, 
which could form a large HII region and surrounding PDR. 
Moreover, the AGN feedback may cause shock heating, which also could contribute to the \cii\ line emissivity.  
In this case, the scenarios of (C) and (E) are further supported. 

Note that if the 
effect of shocks and AGNs is too strong, 
the carbon may be doubly ionized, and then the \cii\ line is rarely emitted. 
Therefore, it is hard to conclude whether the missing treatment of shocks and/or AGNs in current simulations are the major causes of the inadequate \cii\ halo in the simulations.  

It should also be noted that 7 out of 9 sources in the ALMA-HST sample are placed at $5 < z < 6$ when the effect of CMB is weaker than at $z=6.0-7.2$, from which the zoom-in simulation results were taken. 
Because the CMB effect reduces the line luminosity from the diffuse component \citep[e.g.,][]{dacunha2013,zhang2016,pallottini2015,lagache2018}, 
the slight difference in the redshift range may cause the insufficient {\sc [Cii]} line luminosity in the zoom-in simulation results. 

\subsection{Physical Origin of {\sc [Cii]}Halo}

We summarize the possible scenarios 
of what powers the {\sc [Cii]} halo based on the results of Sections \ref{sec:obs}$-$\ref{sec:model}. 
From the observational results, we rule out scenario (A). 
In the zoom-in simulation results, it is hard to conclude which scenario is the most plausible one unless we perform further analyses with different models of SN and AGN feedback, for example.  
The possible scenarios at this point are thus (B), (C), (D), and (E) given the current best estimates of both observational and theoretical results. 

Most importantly, the outflow activities are required in all cases to enrich the CGM with carbon around the normal star-forming galaxies in the early Universe. 
Our results are thus the evidence of outflow remnants in these early star-forming galaxies. 

There are two modes of outflows, hot-mode and cold-mode outflows \citep[e.g.,][]{murray2011,hopkins2014,muratov2015,heckman2017}. 
The hot-mode outflow is defined as the outflow of ionized hydrogen (hot) gas that is heated by supernova (SN) explosions, massive star/AGN radiation. 
Since the cooling time of such hot gas ($\gtrsim 10^{6}$ K) can be longer than the cosmic time at $z\sim$ 5$-$7 ($\sim$1 Gyr; e.g., \citealt{madau2001}), 
it would be difficult to produce the \cii-emitting cold halos from the hot-mode outflow. 
On the other hand, the cold-mode outflow consists of the cold neutral hydrogen gas that is pushed by the radiative and kinetic pressures exerted by SNe, massive stars, and AGNs. 
In this case, the majority of {\sc [Cii]} line emission would be radiated from the PDR in the cold, neutral hydrogen gas clouds. 
Therefore our finding of the {\sc [Cii]} halo 
suggests that outflows in the early star-forming galaxies may be dominated by the cold-mode outflows. 

Since we also find the similarity in the radial surface brightness profiles between the {\sc [Cii]} and Ly$\alpha$ halos (Figure \ref{fig:2comp}), 
the physical origin of the {\sc [Cii]} halo may be related to the Ly$\alpha$ halo. 
Future deep observations of both {\sc [Cii]} and Ly$\alpha$ line emission for individual high-$z$ galaxies are required to comprehensively understand the 
mechanism of the CGM metal enrichment with the theoretical simulations including the radiative transfers of these line emission. 

\section{Summary}
\label{sec:summary}
In this paper, we study the detailed morphology of {\sc [Cii]} line emission via the ALMA visibility-based stacking method 
for normal star-forming galaxies whose {\sc [Cii]} line have been individually detected at $z=5.153-7.142$. 
The visibility-based stacking achieves deep and well-sampled visibility data in the $uv$-plane, 
which enables us to securely investigate the diffuse emission extended over the circum-galactic environment. 
In conjunction with the deep HST/$H$-band data, 
we examine the radial surface brightness profiles of the {\sc [Cii]} line, rest-frame FIR, and UV continuum emission. 
We then discuss the physical origin of the extended {\sc [Cii]} line emission. 
The major findings of this paper are summarized below.
\begin{enumerate}
\item
The visibility-based stacking of our and archival deep ALMA data for 18 galaxies with SFR $\simeq$ 10$-$70$M_\odot$ yr$^{-1}$ at $z=5.153-7.142$ produces 21$\sigma$ and 10$\sigma$ level detections at the peak for the {\sc [Cii]} line and dust continuum emission, respectively. 
The stacked {\sc [Cii]} line morphology is spatially extended more than that of the dust continuum. 
The radial surface brightness profiles of the {\sc [Cii]} line are extended up to a radius of $\sim$10-kpc scale at the 9.2 $\sigma$ level. 
\item
The HST/$H$-band stacking for 9 out of the 18 {\sc [Cii]} line sources that are taken by the deep HST observations 
shows that the radial surface brightness profiles of the {\sc [Cii]} line is significantly extended more than that of the rest-frame UV as well as the rest-frame FIR continuum emission. 
We derive the radial ratio of $L_{\rm [CII]}$/SFR$_{\rm total}$, 
showing that the ratio becomes higher towards the outskirts of halo where the high ratios cannot be explained by the satellite galaxies.  
\item 
The two-component S$\acute{\rm e}$rsic+exponential profile fitting results indicate that the extended {\sc [Cii]} halo component has the scale length of 3.3 $\pm$ 0.1 kpc, which is comparable to the Ly$\alpha$ halo, universally found around the high-$z$ star-forming galaxies. 
In terms of effective radius, the extended {\sc [Cii]} halo component is larger than the central galactic component by a factor of $\sim$5. 
\item 
The state-of-the-art zoom-in cosmological hydrodynamic simulations roughly reproduce the radial surface brightness profile trends of the extended {\sc [Cii]} line emission and the rest-frame FIR, comparable to the rest-frame UV continuum emission. 
{However, the simulations do not reproduce the full extent of the {\sc [Cii]} halo in the outskirts, 
where the simulations might be missing some physical mechanisms associated with the feedback, or still lacking the resolution to resolve the turbulent cascade from large-scale shocks down to the small scales of molecular clouds, if such a process is indeed important for the \cii\ emission in high-$z$ galaxies as \citet{appleton2013} argued. } 
\item 
Although there remain several possible scenarios that can give rise to \cii\ line emission in the CGM, 
the outflow is required in any cases to enrich the primordial CGM with carbon around the early star-forming galaxies. Our results are thus the evidence of outflow remnants in the early star-forming galaxies and suggest that the outflow may be dominated by the cold-mode outflow. 
\end{enumerate}

We thank the anonymous referee for constructive comments and suggestions.
We are grateful to Ivan Marti-Vidal and the Nordic ALMA Regional Center for providing us with helpful CASA software tools and advice on analyzing the data. 
We appreciate Tohru Nagao, Jeremy Blaizot, Peter Mitchell, Takashi Kojima, Shiro Mukae, Yuichi Harikane, Akio Inoue, and Rieko Momose for useful comments and suggestions.  
We are indebted for the support of the staff at the ALMA Regional Center. 
This paper makes use of the following ALMA data: ADS/JAO. ALMA \#2013.1.00815.S, \#2015.1.00834.S, \#2015.1.01111.S, \#2015.1.01105, \#2016.1.01240.S, \#2012.1.00523.S, and \#2012.1.00602.S. 
ALMA is a partnership of the ESO (representing its member states), 
NSF (USA) and NINS (Japan), together with NRC (Canada), MOST and ASIAA (Taiwan), and KASI (Republic of Korea), 
in cooperation with the Republic of Chile. 
The Joint ALMA Observatory is operated by the ESO, AUI/NRAO, and NAOJ. 
This study is supported by World Premier International Research Center Initiative (WPI Initiative), 
MEXT, Japan, and KAKENHI (15H02064, 16J02344, 17H01110, 17H01111, and 17H01114) Grant-in-Aid for Scientific Research (A) 
through Japan Society for the Promotion of Science (JSPS), 
the Grant-in-Aid for JSPS Research Fellow, 
the NAOJ ALMA Scientific Research Grant Number 2017-06B, 
and the Munich Institute for Astro- and Particle Physics (MIAPP) of the DFG cluster of excellence "Origin and Structure of the Universe".
S.F. is supported by the ALMA Japan Research Grant of NAOJ Chile Observatory NAOJ-ALMA-197, 
The 2018 Graduate Research Abroad in Science Program Grant (GRASP2018), 
and the Hayakawa Satio Fund awarded by the Astronomical Society of Japan, 
and NAOJ ALMA Scientific Research Grant Number 2016-01A.
A.F. and R.J.I acknowledge supports from the ERC Advanced Grant INTERSTELLAR H2020/740120 and COSMIC ISM 321302, respectively.

\appendix
\section{Our ALMA Sample} 
\label{sec:appendix}

\begin{figure*}
\begin{center}
\includegraphics[trim=0cm 0cm 0cm 0cm, angle=0,width=1.0\textwidth]{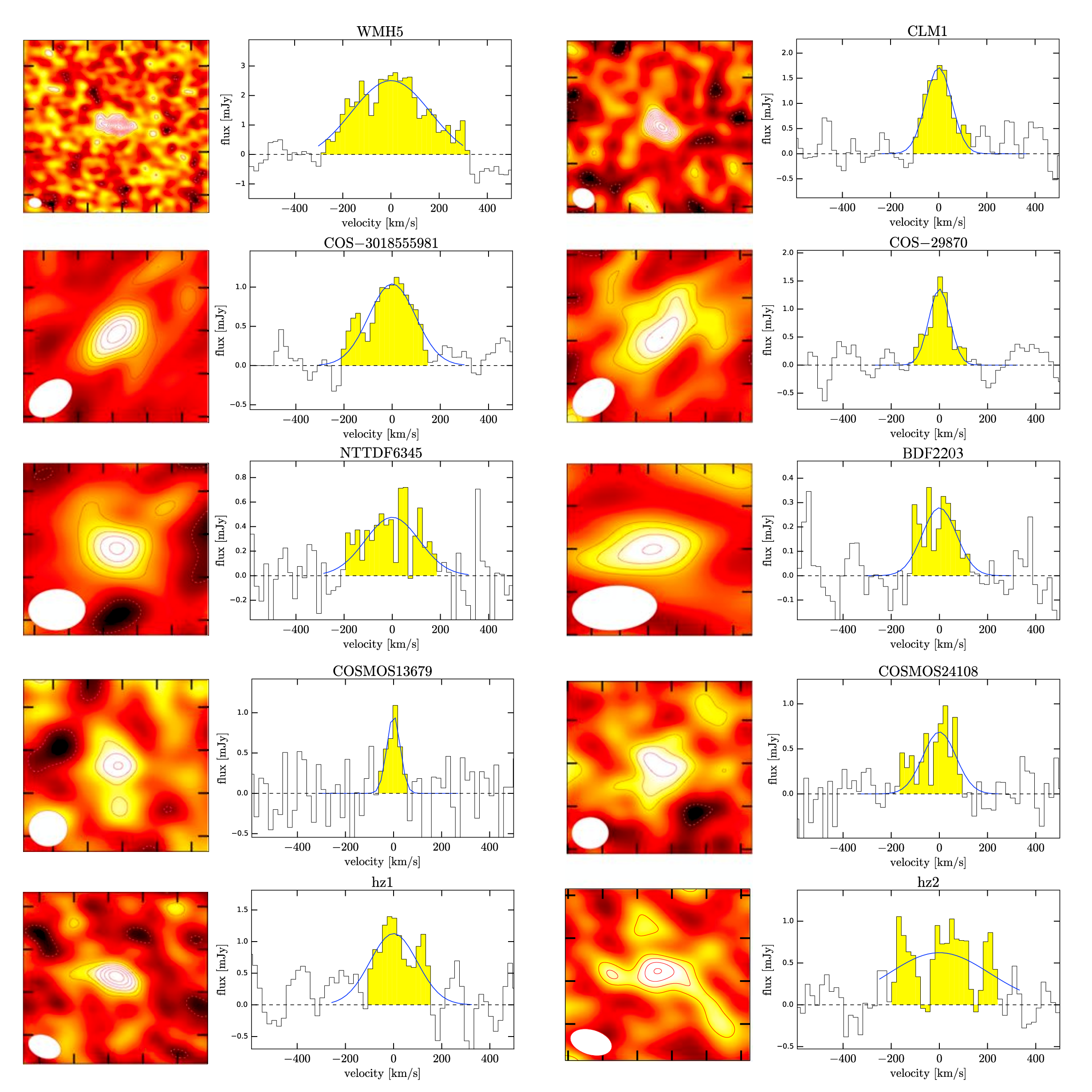}
\caption[]{
The \cii\ line velocity-integrated map (left) and spectrum (right) obtained in our re-analysis of the archival ALMA data 
for the literature sample in Table \ref{tab:our_alma}. 
{\bf Left:} 
Natural-weighted $4''\times4''$ field image of the velocity-integrated \cii\ line intensity (moment zero) 
with contours at the $-2\sigma$ (white), $2\sigma$, $3\sigma$, ..., $10\sigma$ (red) levels. 
The synthesized beam is presented at the bottom left. 
North is up, and east is to the left.
{\bf Right:}
\cii\ line spectrum with an aperture diameter of 1\farcs2. 
The blue curve denotes the best-fit profile of the single Gaussian. 
Here we perform the fitting in the velocity range of $\pm$ 300 km s$^{-1}$ 
from the \cii\ line frequency center estimated in the previous studies.
The yellow shades present the integrated velocity range for the \cii\ line intensity map presented in the left panel. 
The velocities are relative to the center of the best-fit Gaussian. 
\label{fig:postage}}
\end{center}
\end{figure*}

The sources drawn from the literature in our ALMA sample is summarized in Table \ref{tab:our_alma}. 
For this literature sample, Figure \ref{fig:postage} shows the \cii\ line velocity-integrated maps and 
the spectra obtained from our re-analysis of the archival ALMA data. 
We confirm that the spatial morphology and the spectrum shape of the \cii\ lines are consistent with the previous studies
\citep{capak2015,willott2015,pentericci2016,smit2018,carniani2018b}. \\
 
\begin{figure*}
\begin{center}
\includegraphics[trim=0cm 0cm 0cm 0cm, angle=-90,width=1.0\textwidth]{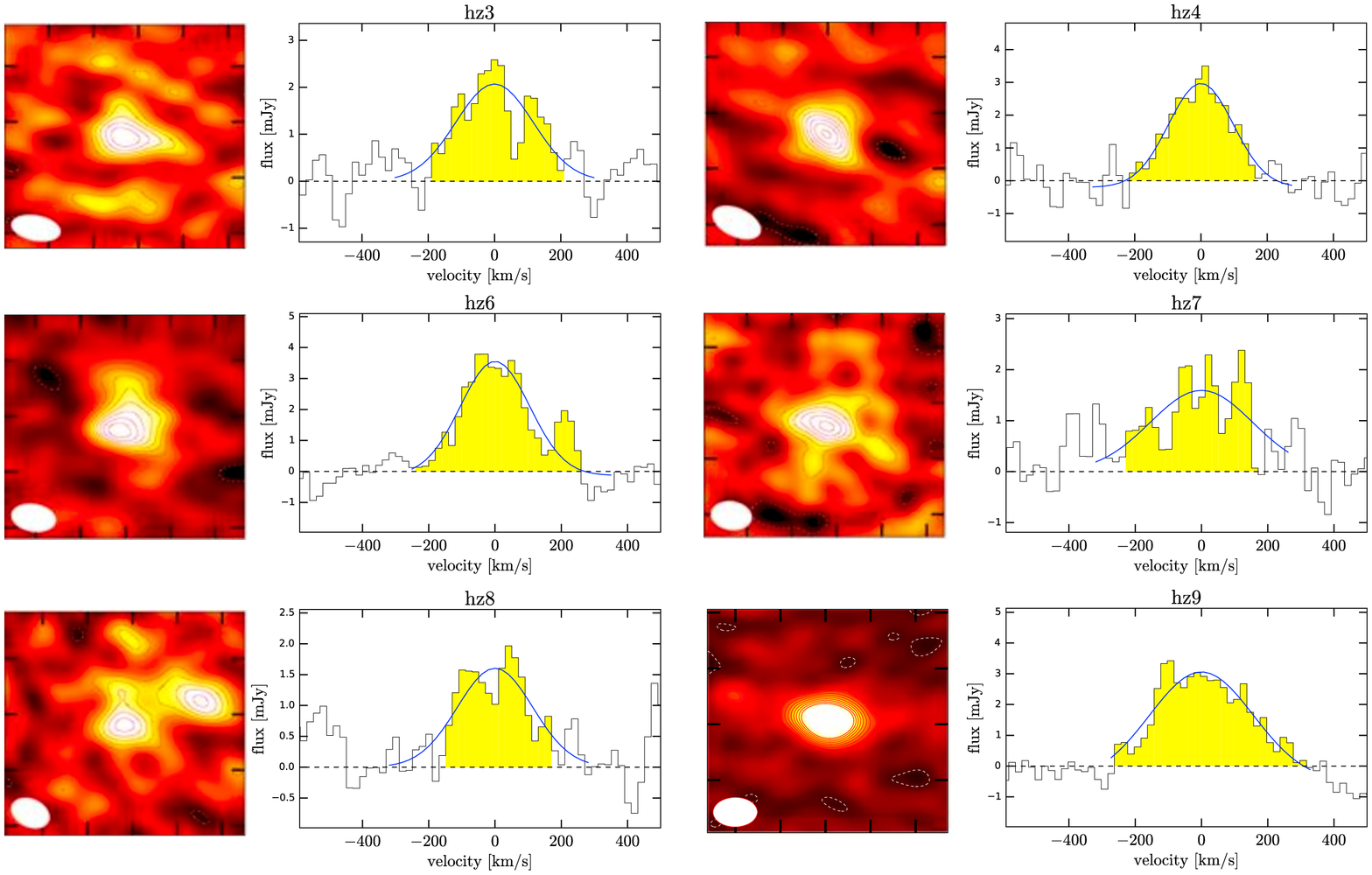}
{\bf Figure \ref{fig:postage}.} (continued)
\end{center}
\end{figure*}

\bibliographystyle{apj}
\bibliography{apj-jour,reference}
\end{document}